\documentclass[a4paper,11pt]{article}
\pdfoutput=1
\usepackage{jheppub}
\usepackage[T1]{fontenc}
\newcommand{\Z}{\mathbb{Z}}
\newcommand{\R}{\mathbb{R}}

\newcommand{\SU}{\mathrm{SU}}
\newcommand{\U}{\mathrm{U}}
\newcommand{\SO}{\mathrm{SO}}

\newcommand{\Gtwo}{\mathrm{G}_2}
\newcommand{\Ffour}{\mathrm{F}_4}
\newcommand{\Eeight}{\mathrm{E}_8}
\newcommand{\GL}{\mathrm{GL}}

\newcommand{\dimension}{{\mathrm{dim}\,}}
\newcommand{\dd}{{\rm{d}}}
\newcommand{\Tr}{{\rm Tr\,}}
\newcommand{\tr}{{\rm Tr\,}}

\newcommand{\dadj}{d_{\mbox{\tiny{a}}}}
\newcommand{\Up}{U_{\mbox{\tiny{p}}}}

\newcommand{\betamin}{\beta_{\mbox{\tiny{min}}}}
\newcommand{\betamax}{\beta_{\mbox{\tiny{max}}}}
\newcommand{\npar}{n_{\mbox{\tiny{par}}}}
\newcommand{\nval}{n_{\mbox{\tiny{val}}}}
\newcommand{\nint}{n_{\mbox{\tiny{int}}}}
\newcommand{\redchisq}{\chi^2_{\tiny\mbox{red}}}
\newcommand{\eq}{\begin{equation}}
\newcommand{\en}{\end{equation}}
\newcommand{\eqar}{\begin{eqnarray}}
\newcommand{\enar}{\end{eqnarray}}

\title{\boldmath Exceptional thermodynamics: The equation of state of $\mathrm{G}_2$ gauge theory}

\author[a]{Mattia Bruno,}
\author[b]{Michele Caselle,}
\author[b,c]{Marco Panero,}
\author[d]{and Roberto Pellegrini}

\affiliation[a]{NIC, DESY\\Platanenallee 6, D-15738 Zeuthen, Germany}
\affiliation[b]{Dipartimento di Fisica, Universit\`a di Torino \& INFN, sezione di Torino\\Via Pietro Giuria 1, I-10125 Torino, Italy}
\affiliation[c]{Instituto de F\'{\i}sica T\'eorica, Universidad Aut\'onoma de Madrid \& CSIC\\Calle Nicol\'as Cabrera 13-15, Cantoblanco E-28049 Madrid, Spain}
\affiliation[d]{Physics Department, Swansea University\\Singleton Park, Swansea SA2 8PP, UK}

\emailAdd{mattia.bruno@desy.de}
\emailAdd{caselle@to.infn.it}
\emailAdd{panero@to.infn.it}
\emailAdd{pyrp@swansea.ac.uk}

\abstract{We present a lattice study of the equation of state in Yang-Mills theory based on the exceptional $\mathrm{G}_2$ gauge group. As is well-known, at zero temperature this theory shares many qualitative features with real-world QCD, including the absence of colored states in the spectrum and dynamical string breaking at large distances. In agreement with previous works, we show that at finite temperature this theory features a first-order deconfining phase transition, whose nature can be studied by a semi-classical computation. We also show that the equilibrium thermodynamic observables in the deconfined phase bear striking quantitative similarities with those found in $\mathrm{SU}(N)$ gauge theories: in particular, these quantities exhibit nearly perfect proportionality to the number of gluon degrees of freedom, and the trace anomaly reveals a characteristic quadratic dependence on the temperature, also observed in $\mathrm{SU}(N)$ Yang-Mills theories (both in four and in three spacetime dimensions). We compare our lattice data with analytical predictions from effective models, and discuss their implications for the deconfinement mechanism and high-temperature properties of strongly interacting, non-supersymmetric gauge theories. Our results give strong evidence for the conjecture that the thermal deconfining transition is governed by a universal mechanism, common to all simple gauge groups.
\vskip10mm
\begin{flushright}
DESY-14-146\\IFT-UAM/CSIC-14-076
\end{flushright} 
}
\begin{document}
\maketitle
\flushbottom

\section{Introduction}
\label{sec:introduction}

Due to its highly non-linear, strongly coupled dynamics, analytical understanding of the strong nuclear interaction remains incomplete~\cite{Brambilla:2014jmp}. Essentially, the fact that the spectrum of physical states is determined by non-perturbative phenomena (confinement and chiral-symmetry breaking) restricts the theoretical toolbox for first-principle investigation of QCD at low energies to numerical simulations on the lattice---while the applicability of weak-coupling expansions is limited to high-energy processes.

At present, one of the major research directions in the study of QCD (both theoretically and experimentally) concerns the behavior of the strong interaction under conditions of finite temperature and/or density. Asymptotic freedom of non-Abelian gauge theories suggests that, at sufficiently high temperatures, ordinary hadrons should turn into a qualitatively different state of matter, characterized by restoration of chiral symmetry and liberation of colored degrees of freedom, which interact with each other through a screened long-range force~\cite{Cabibbo:1975ig}: the quark-gluon plasma (QGP). After nearly twenty years of dedicated experimental searches through relativistic heavy-nuclei collisions, at the turn of the millennium the QGP was eventually discovered at the SPS~\cite{Heinz:2000bk} and RHIC~\cite{Adcox:2004mh, Arsene:2004fa, Back:2004je, Adams:2005dq} facilities.

The measurements performed at RHIC~\cite{Adcox:2004mh, Arsene:2004fa, Back:2004je, Adams:2005dq} and, more recently, at LHC~\cite{Aamodt:2010jd, Aamodt:2010cz, Aamodt:2010pb} reveal a consistent picture: at temperatures $T$ of a few hundreds MeV, QCD is indeed in a deconfined phase, but the QGP behaves as a quite strongly coupled fluid~\cite{Shuryak:2008eq}. These findings are derived from the observation of elliptic flow~\cite{Ackermann:2000tr, Aamodt:2010pa, ATLAS:2011ah, ATLAS:2012at, Chatrchyan:2012ta}, electromagnetic spectra~\cite{Adare:2008ab, Adare:2009qk}, quarkonium melting~\cite{Vogt:1999cu, Gerschel:1998zi, Adare:2006ns, Adare:2008sh, Aad:2010aa, Chatrchyan:2011pe, Chatrchyan:2012np, Abelev:2012rv}, enhanced strangeness production~\cite{Andersen:1998vu, Andersen:1999ym, Antinori:2006ij, Abelev:2006cs, Abelev:2007rw, Abelev:2008zk} and jet quenching~\cite{Aggarwal:2001gn, Adcox:2001jp, Adler:2002xw, Adler:2002tq, Baier:2002tc, Stoecker:2004qu, Antinori:2005cx, Aad:2010bu, Chatrchyan:2011sx, Chatrchyan:2012gt, Spousta:2013aaa, Veres:2013oga}; for a very recent review, see ref.~\cite{Andronic:2014zha}. 

These results indicate that the theoretical investigation of the QGP requires non-perturbative tools, such as computations based on the gauge/string correspondence (whose applications in QCD-like theories at finite temperature are reviewed in refs.~\cite{Gubser:2009md, CasalderreySolana:2011us}) or lattice simulations~\cite{Philipsen:2012nu}. The lattice determination of the deconfinement crossover and chiral transition temperature, as well as of the QGP bulk thermodynamic properties (at vanishing quark chemical potential $\mu$) is settled~\cite{Borsanyi:2013bia, Bazavov:2014pvz, Bhattacharya:2014ara} and accurate results are being obtained also for various parameters describing fluctuations, for the QGP response to strong magnetic fields, et c.~\cite{Szabo:2014iqa}. Due to the Euclidean nature of the lattice formulation, the investigation of phenomena involving Minkowski-time dynamics in the QGP is more challenging, but the past few years have nevertheless witnessed a lot of conceptual and algorithmic advances, both for transport properties~\cite{Meyer:2011gj} and for phenomena like the momentum broadening experienced by hard partons in the QGP~\cite{CaronHuot:2008ni, Majumder:2012sh, Benzke:2012sz, Laine:2012ht, Ghiglieri:2013gia, Laine:2013lia, Laine:2013apa, Panero:2013pla, Cherednikov:2013pba, D'Onofrio:2014qxa}: it is not unrealistic to think that in the near future the results of these non-perturbative calculations could be fully integrated in model computations that provide a phenomenological description of experimentally observed quantities (for a very recent, state-of-the-art example, see ref.~\cite{Burke:2013yra}).

Notwithstanding this significant progress towards more and more accurate numerical predictions, a full theoretical understanding of QCD dynamics at finite temperature is still missing. From a purely conceptual point of view, the problem of strong interactions in a thermal environment can be somewhat simplified, by looking at pure-glue non-Abelian gauge theories. This allows one to disentangle the dynamics related to chiral-symmetry breaking from the problem of confinement and dynamical generation of a mass gap, retaining---at least at a qualitative or semi-quantitative level---most of the interesting features relevant for real-world QCD. As the system is heated up, these theories will interpolate between two distinct limits: one that can be modeled as a gas of massive, non-interacting hadrons (glueballs) at low temperature, and one that is described by a gas of free massless gluons at (infinitely) high temperature. These limits are separated by a finite-temperature region, in which deconfinement takes place. 

In $\SU(N)$ gauge theories, the phenomenon of deconfinement at finite temperature can be interpreted in terms of spontaneous breaking of the well-defined global $\Z_N$ center symmetry~\cite{Polyakov:1975rs} (see also ref.~\cite{Cohen:2014swa} for a very recent work on the subject), and is an actual phase transition: a second-order one for $N=2$ colors, and a discontinuous one for all $N \ge 3$ (see also refs.~\cite{Lucini:2012gg, Panero:2012qx, Lucini:2013qja}). While this is consistent with the interpretation of confinement in non-supersymmetric gauge theories as a phenomenon due to condensation of center vortices~\cite{DelDebbio:1996mh, deForcrand:1999ms} (see also ref.~\cite{Greensite:2003bk} for a discussion), it begs the question, what happens in a theory based on a non-Abelian gauge group with trivial center? In this respect, it is particularly interesting to consider the $\Gtwo$ gauge theory: since this exceptional group is the smallest simply connected group with a trivial center, it is an ideal toy model to be studied on the lattice. For the $\Gtwo$ Yang-Mills theory, even though there is no center symmetry distinguishing the physics at low and at high temperature, one still expects that the physical degrees of freedom at high temperatures be colored ones. Like for $\SU(N)$ theories, this expectation is borne out of asymptotic freedom of the theory, which suggests that the description in terms of a gas of weakly interacting gluons should become accurate when the typical momenta exchanged are large, and this is expected to be the case for a thermal system at high temperature, for which the characteristic energy scale of hard thermal excitations is $O(T)$. Thus, one can still expect that the $\Gtwo$ Yang-Mills theory features a high-temperature regime in which colored states do exist, and define it as the ``deconfined phase'' of the theory.

Note that, although smaller continuous non-Abelian groups with a trivial center do exist, strictly speaking what actually counts is the fundamental group of the compact adjoint Lie group associated with the Lie algebra of the gauge group. For example, the $\SO(3)$ group has a trivial center, but (contrary to some inaccurate, if widespread, claims) this property, by itself, does not make the $\SO(3)$ lattice gauge theory a suitable model for studying confinement without a center,\footnote{Nevertheless, it is worth remarking that the lattice investigation of $\SO(3)$ gauge theory has its own reasons of theoretical interest~\cite{deForcrand:2002vs}.} nor one to be contrasted with the $\SU(2)$ gauge group (which has the same Lie algebra). Indeed, the first homotopy group of $\SO(3)$, which is the compact adjoint Lie group associated with the $B_1$ Lie algebra, is $\Z_2$, i.e. the same as the first homotopy group of the projective special unitary group of degree $2$ (whose associated Lie algebra is $A_1 = B_1$). This leaves only $\Gtwo$, $\Ffour$ and $\Eeight$ as compact simply-connected Lie groups with a trivial center; of these, $\Gtwo$, with rank two and dimension $14$, is the smallest and hence the most suitable for a lattice Monte Carlo study. In fact, numerical simulations of this Yang-Mills theory have already been going on for some years~\cite{Holland:2003jy, Maas:2007af, Liptak:2008gx, Danzer:2008bk, Wellegehausen:2010ai, Wellegehausen:2011sc, Ilgenfritz:2012wg, Greensite:2006sm, Pepe:2006er, Cossu:2007dk, Wellegehausen:2009rq, Bonati:2015uga}. Besides numerical studies, these peculiar features of $\Gtwo$ Yang-Mills theory have also triggered analytical interest~\cite{Poppitz:2012nz, Diakonov:2010qg, Maas:2010qw, Buisseret:2011fq, Deldar:2011fh, Dumitru:2012fw, Lacroix:2012pt, Poppitz:2013zqa, Dumitru:2013xna, Nejad:2014hka, Anber:2014lba, Guo:2014zra}.

Note that the question, whether $\Gtwo$ Yang-Mills theory is a ``confining'' theory or not, depends on the definition of confinement. If one defines confinement as the absence of non-color-singlet states in the physical spectrum, then $\Gtwo$ Yang-Mills theory is, indeed, a confining theory. On the other hand, if one defines confinement as the existence of an asymptotically linear potential between static color sources, then the infrared dynamics of $\Gtwo$ Yang-Mills theory could rather be described as ``screening''. Indeed, previous lattice studies indicate that, at zero and low temperatures, the $\Gtwo$ Yang-Mills theory has a confining phase, in which static color sources in the smallest fundamental irreducible representation $\mathbf{7}$ are confined by string-like objects, up to intermediate distances. At very large distances, however, the potential associated with a pair of fundamental sources gets screened. This is a straightforward consequence of representation theory (and, ultimately, of the lack of a non-trivial $N$-ality for this group): as eq.~(\ref{screening}) in the appendix~\ref{app:G2_properties} shows, the representation $\mathbf{7}$ appears in the decomposition of the product of three adjoint representations $\mathbf{14}$, thus a fundamental $\Gtwo$ quark can be screened by three gluons.

One further reason of interest for a QCD-like lattice theory based on the $\Gtwo$ group is that it is free from the so-called sign problem~\cite{deForcrand:2010ys}: with dynamical fermion fields in the fundamental representation of the gauge group, the introduction of a finite chemical potential $\mu$ does not make the determinant of the Dirac matrix complex, thus the theory can be simulated at finite densities~\cite{Maas:2012wr, Wellegehausen:2013cya}.\footnote{The chiral-symmetry pattern of $\Gtwo$ QCD described in ref.~\cite{Wellegehausen:2013cya} is the following: for $n_f$ massless Dirac flavors at $\mu = 0$, the axial anomaly implies that the theory has an $\SU(2n_f) \otimes \Z_2$ symmetry. In the presence of a quark condensate (or of a finite quark mass $m$), this symmetry gets spontaneously (respectively, explicitly) broken down to $\SO(2n_f) \otimes \Z_2$. Finally, introducing a finite $\mu$ reduces the symmetry down to $\SU(n_f) \otimes \U(1) / \Z_{n_f}$, with the ``baryon'' number associated to the $\U(1)$ factor. Note that, since $\Gtwo$ contains an $\SU(3)$ subgroup, one may wonder if this could open the path to simulating real-world QCD at finite densities without a sign problem, for example by making the six additional $\Gtwo$ gluons not present in the $\SU(3)$ theory arbitrarily heavy, coupling them to a scalar field according to the Brout-Englert-Higgs mechanism~\cite{Englert:1964et, Higgs:1964pj}. This turns out not to be the case: the fundamental representation $\mathbf{7}$ of $\Gtwo$ decomposes into the sum of the trivial ($\mathbf{1}$), the fundamental ($\mathbf{3}$) and the antifundamental ($\bar{\mathbf{3}}$) irreducible representations of $\SU(3)$~\cite{Holland:2003jy}. As a consequence, when viewed in terms of the $\SU(3)$ field content, the chemical potential associated with the $\U(1)$ group described above should be interpreted as an isospin, rather than a baryonic, chemical potential, for which the sign problem is absent~\cite{Son:2000xc}.} As compared to another well-known QCD-like theory which shares this property, namely two-color QCD~\cite{Hands:1999md, Kogut:2001na, Hands:2006ve, Hands:2010gd, Boz:2013rca}, one advantage is that ``baryons'' in $\Gtwo$ QCD are still fermionic states, like in the real world.

Finally, the possibility that gauge theories based on exceptional gauge groups may be relevant in walking technicolor scenarios for spontaneous electro-weak symmetry breaking was studied (via a perturbative analysis) in ref.~\cite{Mojaza:2012zd}.

In this work, we extend previous lattice studies of $\Gtwo$ Yang-Mills theory at finite temperature~\cite{Pepe:2006er, Cossu:2007dk, Wellegehausen:2009rq, Bonati:2015uga} by computing the equation of state in the temperature range $ T \lesssim 3 T_c$, where $T_c$ denotes the critical deconfinement temperature.\footnote{Note that this is the temperature range probed experimentally at the LHC~\cite{Muller:2013dea}, although in real-world QCD deconfinement is a \emph{crossover}, rather than a sharp, first-order transition.} After introducing some basic definitions and the setup of our simulations in section~\ref{sec:setup}, we present our numerical results in section~\ref{sec:results}; then in section~\ref{sec:discussion} we compare them with the predictions of some analytical calculations, pointing out qualitative and quantitative analogies with $\SU(N)$ gauge theories. Finally, in section~\ref{sec:conclusions} we summarize our findings and list possible extensions of the present work. Some general properties of the $\Gtwo$ group and of its algebra are reported in the appendix~\ref{app:G2_properties}.

\section{Setup}
\label{sec:setup}

Our non-perturbative computation of the equation of state in $\Gtwo$ Yang-Mills theory is based on the standard Wilson regularization of the theory on a four-dimensional, Euclidean, hypercubic lattice $\Lambda$ of spacing $a$~\cite{Wilson:1974sk}. Throughout this article, we denote the Euclidean time direction by the index $0$ (or by a subscript ${}_t$), and the spatial directions by $1$, $2$ and $3$ (or by a subscript ${}_s$). Periodic boundary conditions are imposed along the four directions. Using natural units $c=\hbar=k_{\tiny{\mbox{B}}}=1$, the physical temperature is given by the inverse of the length of the shortest side of the system (which we take to be in the direction $0$), $T=1/(aN_t)$, while the other three sides of the hypertorus have equal lengths, denoted by $L_s = a N_s$. In order to avoid systematic uncertainties caused by finite-volume effects, we always take $ L_s T \gg 1$; in practice, previous studies of $\SU(N)$ Yang-Mills thermodynamics have shown that, at the temperatures of interest for this work, finite-volume effects are negligible for $L_s T \gtrsim 4$~\cite{Gliozzi:2007jh, Panero:2008mg, Mykkanen:2012ri}. 

To extract vacuum expectation values at low temperature (in the confining phase), we carry out simulations on lattices of sizes $L_s^4$: for the parameters of our simulations, this choice corresponds to temperatures which are sufficiently ``deep'' in the confining phase---meaning temperatures, at which the values of the bulk thermodynamic quantities, that we are interested in, are well below the statistical precision of our data.

The partition function of the lattice system is defined by the multiple group integral
\eq
\label{lattice_partition_function}
Z = \int \prod_{x \in \Lambda} \prod_{\alpha=0}^3 {\dd}U_\alpha(x) \, e^{-S_{\mbox{\tiny{W}}}},
\en
where ${\dd}U_\alpha(x)$ denotes the Haar measure for the generic $U_\alpha(x)$ matrix, which represents the parallel transporter on the oriented bond from $x$ to $x+a\hat{\alpha}$. The $U_\alpha(x)$ matrices take values in the representation of the $\Gtwo$ group in terms of real $7 \times 7$ matrices,\footnote{Actually, for part of our simulations we also used a different algorithm, using complex matrices and based on the decomposition of the $\Gtwo$ group discussed in ref.~\cite{Macfarlane:2002hr}---see ref.~\cite{Pepe:2006er} for details. Although, intuitively, one would expect the implementation based on the representation in terms of real matrices to be faster to simulate than the one involving complex matrices, this is not necessarily the case. What really determines the efficiency of the simulation algorithm is not simply the CPU time necessary to multiply gauge link variables with each other, but rather the time necessary to sufficiently decorrelate the degrees of freedom in a sequence of configurations produced in a Markov chain during the Monte~Carlo process. In this respect, the implementation in terms of complex matrices seems to be more efficient than the one based on real ones, offsetting the drawback of a larger number of elementary multiplications required for products of complex factors. However, clearly this efficiency gain depends (strongly) on the dynamics: for example, in the strong-coupling limit, in which the weight of the configurations contributing to eq.~(\ref{lattice_partition_function}) reduces to the product of the Haar measures for the link variables, it is trivial to obtain a sequence of decorrelated configurations, simply by choosing the new $\Gtwo$ elements randomly in the group. While we stress that we did not carry out a systematic performance study to compare the efficiency of our two algorithms in different regimes, we remark that we simply observed that, at least in the regime that we investigated, the implementation based on complex matrices is not less efficient than the one based on real matrices.} while
\eq
\label{Wilson_lattice_gauge_action}
S_{\mbox{\tiny{W}}}= - \frac{1}{g^2} \sum_{x \in \Lambda} \sum_{0 \le \mu < \nu \le 3} \Tr U_{\mu,\nu}(x)
\en
is the gauge-invariant Wilson lattice action~\cite{Wilson:1974sk}. Here, $g$ is the bare lattice coupling and
\eq
\label{plaquette}
U_{\mu,\nu}(x) = U_\mu(x) U_\nu(x+a\hat\mu) U^\dagger_\mu(x+a\hat\nu) U^\dagger_\nu(x)
\en
denotes the plaquette stemming from site $x$ and lying in the oriented $(\mu,\nu)$ plane. In the following, we also introduce the Wilson action parameter $\beta$, which for this theory can be defined as $\beta=7/g^2$.

As usual, expectation values of gauge-invariant quantities $\mathcal{O}$ are then defined as
\eq
\label{vev}
\langle \mathcal{O} \rangle = \frac{1}{Z} \int \prod_{x \in \Lambda} \prod_{\alpha=0}^3 {\dd}U_\alpha(x) \, \mathcal{O} \, e^{-S_{\mbox{\tiny{W}}}}
\en
and can be computed numerically, via Monte Carlo integration. To this purpose, we generated ensembles of matrix configurations using an algorithm that performs first a heat-bath update, followed by five to ten overrelaxation steps, on an $\SU(3)$ subgroup of $\Gtwo$ (in turn, both the heat-bath and overrelaxation steps are based on three updates of $\SU(2)$ subgroups~\cite{Cabibbo:1982zn}). Finally, a $\Gtwo$ transformation is applied, in order to ensure ergodicity. The parametrization of $\Gtwo$ that we used is described in refs.~\cite{Cacciatori_mathph0503054, Cacciatori:2005yb}. After a thermalization transient, we generate the ensemble to be analyzed by discarding a certain number (which depends on the physical parameters of the simulation---in particular, on the proximity to the deconfinement temperature, which affects the autocorrelation time of the system) of intermediate configurations between those to be used for our analysis. Typically, the number of configurations for each combination of parameters ($\beta$, $N_s$ and $N_t$) is $O(10^4)$. This leads to an ensemble of (approximately) statistically independent configurations, allowing us to bypass the problem of coping with difficult-to-quantify systematic uncertainties due to autocorrelations. Throughout this work, all statistical errorbars are computed using the gamma method~\cite{Wolff:2003sm}; a comparison on a data subset shows that the jackknife procedure~\cite{bootstrap_jackknife_book} gives roughly equivalent results.

We computed expectation values of hypervolume-averaged, traced Wilson loops at zero temperature
\eq
\label{Wilson_loop}
W(r,L) = \frac{1}{6N_s^4} \sum_{x \in \Lambda} \sum_{0 \le \mu < \nu \le 3} \frac{1}{7} \tr \left\{ \mathcal{L}_\mu^r(x) \mathcal{L}_\nu^L(x+r\hat{\mu}) \left[ \mathcal{L}_\mu^r(x+L\hat{\nu}) \right]^\dagger \left[ \mathcal{L}_\nu^L(x) \right]^\dagger \right\},
\en
with
\eq
\label{lines}
\mathcal{L}_\mu^r(x) = \prod_{n=0}^{r/a-1} U_\mu \left( x + n a \hat\mu \right),
\en
as well as of volume-averaged, traced Polyakov loops at finite temperature
\eq
\label{Polakov_loop}
P = \frac{1}{N_s^3} \sum_{x \in V_{t=0}} \frac{1}{7} \tr \mathcal{L}_0^{1/T}(x)
\en
(where $V_{t=0}$ denotes the spatial time-slice of $\Lambda$ at $t=0$) and of hypervolume-averaged plaquettes (both at zero and at finite temperature).

From the expectation values of Wilson loops, which we computed with the multilevel algorithm~\cite{Luscher:2001up, Kratochvila:2003zj}, the heavy-quark potential $V(r)$ can be extracted via
\eq
\label{string_tension}
V(r) = - \frac{1}{L} \ln \langle W(r,L) \rangle.
\en

In practice, since we are bound to use loops of finite sizes, in order to avoid possible contamination from an excited state, we perform both a single- ($k_1=0$) and a two-state ($k_1$ free) fit
\eq
\label{two_state_fit}
\langle W(r,L) \rangle = e^{-L V(r)} + k_1 e^{-L V_1(r)},
\en
extracting our results for $V$ from fits in the $L$ range where the results including or neglecting the second addend on the right-hand side of eq.~(\ref{two_state_fit}) are consistent, within their uncertainties.

We determined the values of the lattice spacing $a$ in our lattice simulations (as a function of $\beta$) non-perturbatively, comparing different methods. A common strategy to set the scale in lattice simulations of $\SU(N)$ Yang-Mills theories is based on the extraction of the value of the string tension in lattice units, $\sigma a^2$, which can be obtained from a two-parameter fit of the static quark-antiquark potential $V(r)$ to the Cornell form
\eq
\label{Cornell}
V(r) = \sigma r + V_0 - \frac{\pi}{12r}.
\en
Strictly speaking, eq.~(\ref{Cornell}) is not an appropriate functional form to model the potential in $\Gtwo$ Yang-Mills theory at large $r$, because this theory is screening in the infrared limit. However, one can nevertheless use it as an approximate description of the potential at the distances probed in this work, because gluon screening sets in at much longer distances (see also ref.~\cite{Greensite:2006sm}). This enables one to extract the values of the string tension in lattice units $\sigma a^2$, at each value of $\beta$. Note that the coefficient of the $1/r$ term in eq.~(\ref{Cornell}) is uniquely fixed by the central charge of the underlying low-energy effective theory describing transverse fluctuations of the confining string in four spacetime dimensions~\cite{Luscher:1980fr}, while we neglect possible higher-order (in $1/r$) terms~\cite{Aharony:2013ipa}.

An alternative method to set the scale, which is more appropriate for intermediate distances (and conceptually better-suited for an asymptotically screening theory), was introduced in ref.~\cite{Sommer:1993ce} (see also ref.~\cite{Necco:2001xg} for a high-precision application in $\SU(3)$ Yang-Mills theory) and is based on the computation of the quark-antiquark force $F(r)$. Using the force, one can introduce the length scales $r_0$ and $r_1$, respectively defined by
\begin{equation}
\label{r_0_definition}
r_0^2 F(r_0)=1.65
\end{equation}
and
\begin{equation}
\label{r_1_definition}
r_1^2 F(r_1)=1.
\end{equation}
The physical values of these scales (in QCD with dynamical quarks) are $r_0=0.472(5)$~fm ~\cite{Aoki:2009ix, Davies:2009tsa, Bazavov:2011nk, Arthur:2012opa} and $r_1=0.312(3)$~fm~\cite{Follana:2007uv, Davies:2009tsa, Bazavov:2011nk, Arthur:2012opa, Dowdall:2013rya}. Note that, as compared to the method based on the string tension, setting the scale using $r_0$ or $r_1$ also has the practical advantage that it is not necessary to go to the large-distance limit, and an improved definition of the force allows one to reduce discretization effects. For these reasons, we used $r_1$ to set the scale in our lattice simulations.

Other methods to set the scale are discussed in refs.~\cite{Sommer:2014mea, Luscher:2010iy} and in the works mentioned therein.

The phase structure of the lattice theory is revealed by the expectation value of the plaquette---averaged over the lattice hypervolume and over the six independent $(\mu,\nu)$ planes---at $T=0$, that we denote as $\langle \Up \rangle_0$: similarly to what happens in $\SU(N)$ gauge theories, as $\beta$ is increased from zero to large values, $\langle \Up \rangle_0$ interpolates between a strong-coupling regime, dominated by discretization effects, and a weak-coupling regime, analytically connected to the continuum limit. The two regions are separated by a rapid crossover (or, possibly, a first-order transition) taking place at $\beta_c \simeq 9.45$~\cite{Pepe:2006er, Cossu:2007dk}. This bulk transition is unphysical, and, in order to extract physical results for the thermodynamics of the theory, all of our simulations are performed in the region of ``weak'' couplings, $\beta > \beta_c$, which is connected to the regime of continuum physics.

In the $\beta > \beta_c$ region, the finite-temperature deconfinement transition is probed by studying the distribution of values and the Monte Carlo history of the bare Polyakov loop (after thermalization): in the confining phase, the distribution of $P$ is peaked near zero, whereas a well-defined peak at a finite value of $P$ develops, when the system is above the critical deconfinement temperature $T_c$. In the vicinity of $T_c$, both the distribution of $P$ values $\rho(P)$ (with two local maxima, separated by a region in which $\rho(P)$ gets suppressed when the physical volume of the lattice is increased) and the typical Monte Carlo histories of $P$ (featuring tunneling events, which become more and more infrequent when the lattice volume is increased, between the most typical values of $P$) give strong indication that the deconfining transition is of first order, in agreement with previous studies~\cite{Pepe:2006er, Cossu:2007dk}.\footnote{Note that, since $\Gtwo$ Yang-Mills theory lacks an underlying exact center symmetry, it is possible that this deconfinement transition could be turned into a crossover by suitably deforming the theory (for example through the inclusion of additional fields in the Lagrangian).}

The equilibrium thermodynamics quantities of interest in the present work are the pressure $p$, the energy density per unit volume $\epsilon$, the trace of the energy-momentum tensor $\Delta$ (which has the meaning of a \emph{trace anomaly}, being related to the breaking of conformal invariance of the classical theory by quantum effects), and the entropy density per unit volume $s$: they can be obtained from the finite-temperature partition function $Z$ via
\eqar
\label{p_definition} p & = & T \left. \frac{\partial \ln Z}{\partial V} \right|_T , \\
\label{Delta_definition} \epsilon & = & \frac{T^2}{V} \left. \frac{\partial \ln Z}{\partial T} \right|_V , \\
\label{Delta_epsilon_minus_3p} \Delta & = & \epsilon - 3p , \\
\label{s_definition} s & = &  \frac{1}{V} \ln Z +\frac{\epsilon}{T}.
\enar
Introducing also the free-energy density
\eq
\label{f_definition}
f = - \frac{T}{V} \ln Z ,
\en
the pressure can be readily computed using the $p=-f$ identity, which holds in the thermodynamic limit. Using the standard ``integral method'' of ref.~\cite{Engels:1990vr}, $p$ (or, more precisely, the difference between the pressure at finite and at zero temperature) can thus be obtained as
\eq
\label{integral_method}
p = \frac{T}{V} \int_{\beta_0}^{\beta} d \beta^\prime \frac{\partial \ln Z}{\partial \beta^\prime} = 6 T^4 \int_{\beta_0}^{\beta} d \beta^\prime \left( \langle \Up \rangle_T - \langle \Up \rangle_0 \right),
\en
where $\langle \Up \rangle_T$ denotes the plaquette expectation value at the temperature $T$, and $\beta_0$ corresponds to a point sufficiently deep in the confining phase, i.e. to a temperature at which the difference between $p$ and its zero-temperature value is negligible.  The integral in the rightmost term of eq.~(\ref{integral_method}) is computed numerically, by carrying out simulations at a set of (finely spaced) $\beta$ values within the desired integration range, and performing the numerical integration according to the trapezoid rule. Although more sophisticated methods, like those described in ref.~\cite[appendix A]{Caselle:2007yc}, could allow us to reduce the systematic uncertainties related to the numerical integration, it turns out that this would have hardly any impact on the total error budget of our results.

The lattice determination of the trace of the energy-momentum tensor $\Delta$ (in units of $T^4$) is even more straightforward, as
\eq
\label{lattice_rescaled_trace}
\frac{\Delta}{T^4} = 6 N_t^4 
\left( \langle \Up \rangle_0 - \langle \Up \rangle_T \right)
\cdot a \frac{\partial \beta}{\partial a},
\en
but it requires an accurate determination of the relation between $\beta$ and $a$. We carried out the latter non-perturbatively, using the $r_1$ scale extracted from $F(r)$, as discussed above.\footnote{Note that eq.~(\ref{lattice_rescaled_trace}) reveals one technical challenge in this computation: on the one hand, as we mentioned above, the simulations have to be carried out at values of the coupling in the region analytically connected to the continuum limit, i.e. $\beta > \beta_c$. In practice, this corresponds to relatively fine lattice spacings, or $N_t \gtrsim 5$. On the other hand, the $N_t^4$ factor appearing on the right-hand side of eq.~(\ref{lattice_rescaled_trace}) shows that the physical signal is encoded in a difference of average plaquette values $\langle \Up \rangle_0$ and $\langle \Up \rangle_T$ (both of which remain finite), that becomes increasingly small when the continuum limit is approached. In practice, the computational costs due to this technical aspect severely restrict the range of $N_t$ values which can be used, and, as a consequence, the lever arm to control the continuum limit.}

Finally, $\epsilon$ and $s$ can be readily computed as linear combinations of $p$ and $\Delta$, using eq.~(\ref{Delta_epsilon_minus_3p}) and the thermodynamic identity
\eq
\label{lattice_entropy_density}
sT = \Delta + 4p.
\en

It is worth noting that alternative methods to determine the equation of state have been recently proposed in ref.~\cite{Umeda:2008bd} and in refs.~\cite{Giusti:2010bb, Giusti:2011kt, Giusti:2012yj, Giusti:2014ila}.

\section{Numerical results}
\label{sec:results}

The first set of numerical results that we present in this section are those aimed at the non-perturbative determination of the scale, i.e. of the relation between the parameter $\beta$ and the corresponding lattice spacing $a$. As explained in sect.~\ref{sec:setup}, we compared two different methods to extract this relation: first by evaluating the string tension in lattice units from the area-law decay of large Wilson loops at zero temperature, and then by computing the force and evaluating the $r_1$ scale defined in eq.~(\ref{r_1_definition}) in lattice units.

\begin{figure}
\begin{center}
\includegraphics*[width=\textwidth]{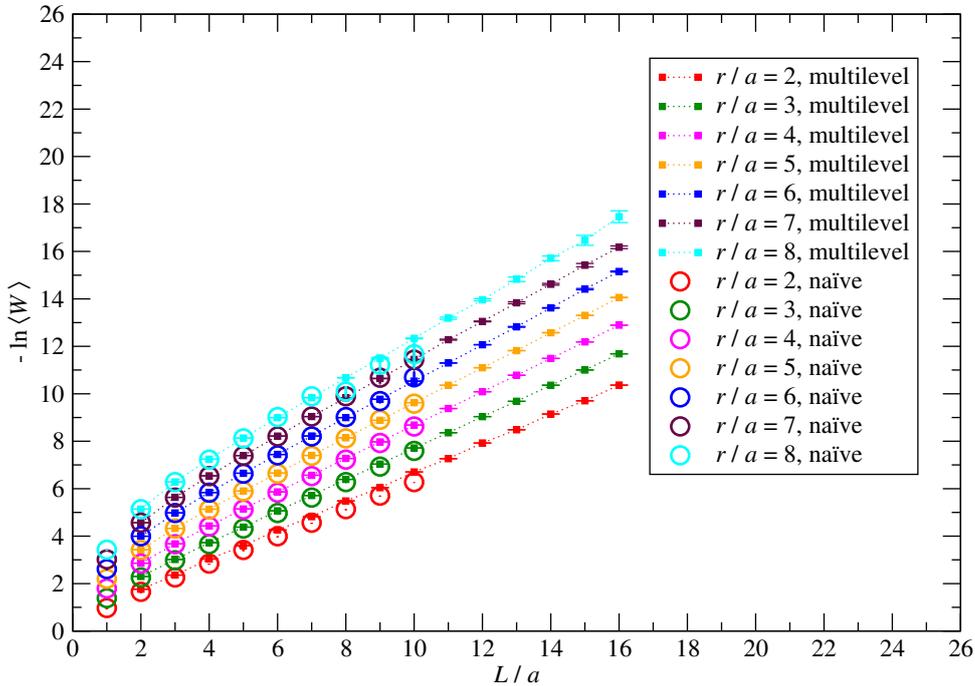}
\caption{Comparison of Wilson loop expectation values (in logarithmic scale), as a function of the loop sizes, computed with and without the multilevel algorithm. The 
plot displays a sample of our results at a fixed lattice spacing, from simulations at $\beta=10.4$.
}
\label{fig:Wilson_loops}
\end{center}
\end{figure}

To achieve high precision, the numerical computation of Wilson loops at zero temperature was carried out using the multilevel algorithm~\cite{Luscher:2001up, Kratochvila:2003zj}, which yields exponential enhancement of the signal-to-noise ratio for long loops. Fig.~\ref{fig:Wilson_loops} shows results for the opposite of the logarithm of Wilson loops of different widths $r/a$ (symbols of different colors) as a function of the loop length in lattice units $L/a$. The plot displays the results from our simulations at $\beta=10.4$. The comparison of results obtained from a na\"{\i}ve, brute-force computation (empty circles) and with our implementation of the multilevel algorithm (filled squares) clearly shows that the latter are in complete agreement with the former for short loops, and that the multilevel algorithm outperforms the brute-force approach for large values of $L$, where the numerical values obtained with the latter are affected by dramatic loss of relative precision.

Our results for the string tension in lattice units, as a function of $\beta$, are reported in table~\ref{tab:string_tension}, which also shows the sizes of the lattices on which the corresponding simulations were carried out (all these zero-temperature simulations were performed on hypercubic lattices of size $L_s^4$), the range of $r/a$ values used in the fit (including both extrema), as well as the results for the constant term in the Cornell potential in lattice units and the reduced $\chi^2$. Note that these are two-parameter fits to eq.~(\ref{Cornell}), while, as mentioned in section~\ref{sec:setup}, the $1/r$ term is fixed to be the L\"uscher term~\cite{Luscher:1980fr}. In principle, for data at small $r/a$ one could use an improved definition of the lattice distance~\cite{Luscher:1995zz, Necco:2001xg}, however this type of correction becomes rapidly negligible at large distances, and hence 
should not change significantly our estimates of $\sigma a^2$, which are dominated by infrared physics.

\begin{table}
\begin{center}
\begin{tabular}{|c|c|c|c|c|c|}
\hline
 $\beta$ & $L_s/a$ & $r/a$ range & $\sigma a^2$ & $a V_0$ & $\redchisq$ \\
\hline
 $  9.6   $ & $32 $ & $[4:7]$ & $0.1335(77)$  & $0.740(35)$  & $0.4$  \\
 $  9.8   $ & $32 $ & $[4:8]$ & $0.0715(17)$  & $0.7469(78)$ & $1.7$  \\
 $ 10.0   $ & $32 $ & $[4:8]$ & $0.0471(11)$  & $0.7191(46)$ & $0.12$ \\
 $ 10.2   $ & $32 $ & $[4:7]$ & $0.03189(77)$ & $0.6697(35)$ & $1.04$ \\
 $ 10.4   $ & $32 $ & $[4:8]$ & $0.02369(66)$ & $0.6729(30)$ & $0.87$ \\
 $ 10.8   $ & $32 $ & $[4:9]$ & $0.01682(58)$ & $0.6546(25)$ & $1.82$ \\
\hline
\end{tabular}
\end{center}
\caption{Summary of the fits of our Monte Carlo results for the static quark-antiquark potential, as a function of $r$, to the Cornell form eq.~(\protect\ref{Cornell}), for the string tension $\sigma$ and the constant term $V_0$, both in lattice units. The numerical results for $V$ are obtained from the expectation values of large Wilson loops computed at zero temperature on lattices of size $L_s^4$ using the multilevel algorithm, according to the procedure discussed in sect.~\ref{sec:setup}. The interval of distances $r$ included in the fits and the corresponding reduced-$\chi^2$ values are also shown.}
\label{tab:string_tension}
\end{table}

\begin{figure}
\begin{center}
\includegraphics*[width=0.8\textwidth]{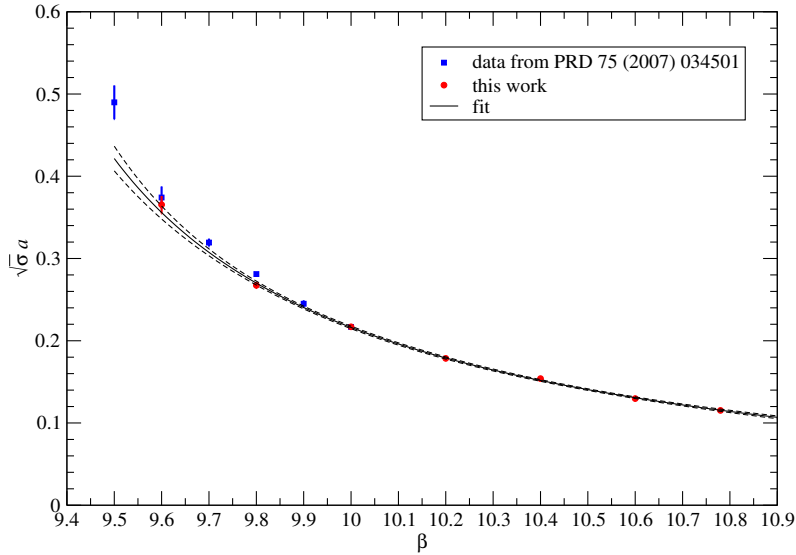}
\caption{Results for the square root of the string tension in lattice units, as a function of the Wilson-action parameter $\beta$. The plot shows our simulation results (red circles) in comparison with those from ref.~\cite{Greensite:2006sm} (blue squares), as well as the interpolating function described by eq.~(\protect\ref{a_interpolation}) for the values of the parameters listed in the text (solid black curve), with the associated uncertainty (dashed black curves).\label{fig:a_interpolation}}
\end{center}
\end{figure}

The values of $\sigma a^2$ thus computed non-perturbatively can be interpolated by a fit to a suitable functional form, in order to get an expression for $a$ as a function of $\beta$ in the region of interest. In principle, this can be done in various ways (see, for example, refs.~\cite{Allton:1996dn, Necco:2001xg, Allton:2008ty}), which, in particular, can include slightly different parametrizations of the discretization effects. One of the simplest possibilities is to fit our data for the logarithm of $\sigma a^2$ to a polynomial of degree $\npar-1$ in $(\beta - \beta_0)$, where $\beta_0 $ is a value within the range of simulated data. Choosing $\beta_0 = 10.2$, a parabolic fit with three parameters, however, yields a large $\redchisq \simeq 6.4$. Different choices of $\beta_0$ and/or of $\npar \ge 3$ give interpolating functions that are only marginally different (within the uncertainties of the fitted parameters) and do not bring $\redchisq$ down to values close to $1$. For example, using a cubic, rather than  quadratic, polynomial, the fitted curve changes slightly, and the $\redchisq$ (using all points) changes from $6.4$ to $5.2$. These unsatisfactory results hint at discretization effects---an indication confirmed by that fact that, excluding the data corresponding to the coarsest lattice spacing, at $\beta=9.6$, from the fit, the $\redchisq$ goes down to $2.2$---and call for a modeling of our lattice results via a functional form that could (at least partially) account for lattice-cutoff systematics. Therefore, following ref.~\cite{Allton:1996dn}, we chose to interpolate the values for the string tension in lattice units computed non-perturbatively by a fit to
\begin{equation}
\label{a_interpolation}
\sqrt{\sigma} a = \frac{ c_1 f(\beta)}{1 + c_3 f^2(\beta)}, \qquad f(\beta) = e^{-c_2 (\beta - \beta_0)},
\end{equation}
with $\beta_0 = 9.9$. This yields $c_1=0.139(25)$, $c_2=0.450(99)$ and $c_3=-0.42(10)$, with $\redchisq=0.73$. Among the systematic uncertainties affecting this scale setting are, for example, those related to the possibility of adding a $c_4 f^4(\beta)$ term in the denominator, or modifying the functional form for $f$ (e.g. multiplying it by a polynomial in $\beta$): while not necessarily better-motivated from a theoretical point of view,\footnote{In particular, the knowledge of terms predicted by one- or two-loop weak-coupling expansions is of little guidance in this range of couplings, far from the perturbative regime.} these alternative parametrizations do not lead to significant changes in the determination of the scale. The curve obtained from the fit to  eq.~(\ref{a_interpolation}) is shown in fig.~\ref{fig:a_interpolation}, together with the data; in addition to our results, we also show those obtained in ref.~\cite{Greensite:2006sm}, which are essentially compatible with our interpolation (within its uncertainty).

From the results of our three-parameter fit to eq.~(\ref{a_interpolation}), the lattice spacing $a$ could be determined at any $\beta$ value within the range of interpolation---provided one defines a physical value for the string tension $\sigma$: to make contact with real-world QCD, one can, for example, set $\sigma=(440$~MeV$)^2$. In addition, the derivative of $\ln \left( \sqrt{\sigma} a \right)$ with respect to $\beta$ is also obtained as
\begin{equation}
\label{derivative_of_ln_a}
- c_2 \frac{1-c_3 f^2(\beta)}{1+c_3 f^2(\beta)}
\end{equation}
and the inverse of this quantity could be used in the computation of the trace of the energy-momentum tensor $\Delta$, according to eq.~(\ref{lattice_rescaled_trace}).

\begin{figure}
\begin{center}
\includegraphics*[width=0.8\textwidth]{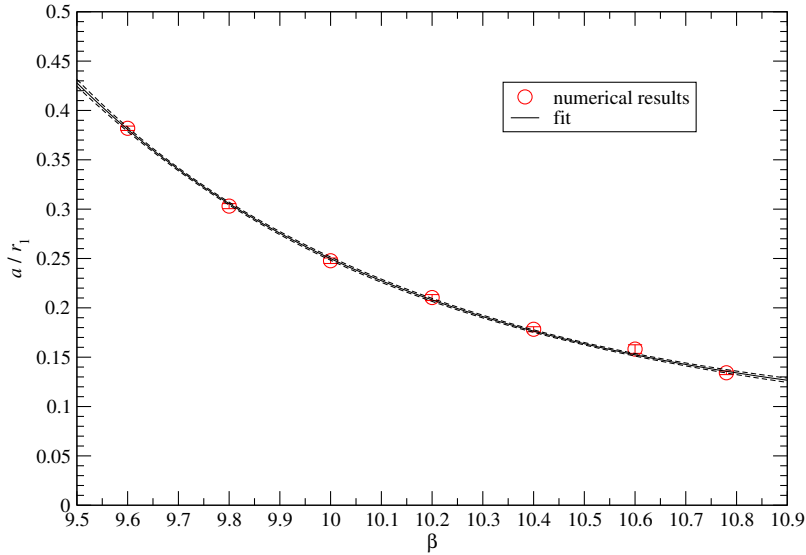}
\caption{Results for the ratio of the lattice spacing over the $r_1$ scale, defined by eq.~(\ref{r_1_definition}) in ref.~\cite{Sommer:1993ce}, as a function of the Wilson action parameter $\beta$. The red symbols denote our numerical results, while the solid black curve and the dashed black lines indicate their fit to eq.~(\ref{r_1_fit}) and the corresponding uncertainty.\label{fig:r1}}
\end{center}
\end{figure}

However, as we mentioned in sect.~\ref{sec:setup}, a determination of the scale based on the extraction of $r_0$ or $r_1$ is better-suited for this theory. Thus we proceeded to evaluate the lattice spacing in units of $r_1$, using the techniques described in ref.~\cite{Necco:2001xg} (including, in particular, the tree-level improved definition of the lattice force introduced in ref.~\cite{Luscher:1995zz}). The results are shown in fig.~\ref{fig:r1}, together with their fit to
\begin{equation}
\label{r_1_fit}
\ln \left( \frac{a}{r_1} \right) = \sum_{j=0}^2 d_j (\beta-\beta_1)^j,
\end{equation}
with $\beta_1=10.1$. The fitted parameters are $d_0=-1.4807(71)$, $d_1=-0.916(10)$, $d_2=0.230(30)$ and the reduced $\chi^2$ equals $1.37$. Our results for $r_1/a$ are reported in table~\ref{tab:r1}.

\begin{table}
\begin{center}
\begin{tabular}{|c|c|}
\hline
 $\beta$ & $r_1/a$ \\
\hline
 $  9.6   $ & $ 2.618(15)  $ \\
 $  9.8   $ & $ 3.299(28)  $ \\
 $ 10.0   $ & $ 4.037(46)  $ \\
 $ 10.2   $ & $ 4.752(68)  $ \\
 $ 10.4   $ & $ 5.609(81)  $ \\
 $ 10.6   $ & $ 6.32(17)   $ \\
 $ 10.78  $ & $ 7.45(10)   $ \\
\hline
\end{tabular}
\end{center}
\caption{Results for $r_1/a$ obtained from our lattice computations at different values of $\beta$.}
\label{tab:r1}
\end{table}

Note that the scale setting in terms of the $r_1$ parameter also appears to be ``cleaner'' than the one based on the string tension, that we discussed above, and insensitive to discretization effects, within the precision of our data. For these reasons, we decided to use $r_1$ to set the scale in our simulations.

Next, we proceeded to simulations at finite temperature, which we carried out on lattices of sizes $N_s^3 \times N_t$ (in units of the lattice spacing), where the shortest size $N_t$ defines the temperature via $T = 1 / ( a N_t )$, while $N_s \gtrsim 4 N_t$. As we already pointed out, by virtue of thermal screening, an aspect ratio of the order of $4$ (or larger) for the ``temporal'' cross-section of the system is known to provide a sufficient suppression of finite-volume effects in $\SU(N)$ gauge theories at the temperatures under consideration~\cite{Panero:2008mg}, while sizable corrections to the thermodynamic quantities are expected to appear at much higher temperatures~\cite{Gliozzi:2007jh}. Some tests on lattices of different spatial volume confirm that this is the case for $\Gtwo$ Yang-Mills theory, too, and did not give us any evidence of significant finite-volume corrections. 

The parameters of our simulations are summarized in table~\ref{tab:finite_T_parameters}. In order to compute the pressure with respect to its value at a temperature close to zero, according to the method described in sect.~\ref{sec:setup}, for each set of finite-temperature simulations we also carried out Monte Carlo simulations on lattices of sizes $(aN_s)^4$, at the same values of the lattice spacing.

\begin{table}
\begin{center}
\begin{tabular}{| c | c | c | c | c |}
\hline
 $N_s$  & $N_t$ & $n_\beta$ & $\beta$-range \\
\hline 
 $ 16 $ &  $5$  &  $35$  & $[9.62,10.64]$ \\
 $ 32 $ &  $6$  &  $60$  & $[9.6,10.78]$  \\
 $ 32 $ &  $8$  &  $30$  & $[9.8,10.7]$   \\
\hline
\end{tabular}
\end{center}
\caption{Parameters of the finite-temperature lattice calculations carried out in this work: $N_s$ and $N_t$ denote the lattice sizes along the space-like and time-like directions (in units of the lattice spacing), $n_\beta$ denotes the number of $\beta$ values simulated, for $\betamin \le \beta \le \betamax$. All simulations at finite temperature are carried out on lattices of sizes $(aN_s)^3 \times (aN_t)$, while those deep in the confined phase (i.e., approximately at $T=0$) are performed on lattices of sizes $(aN_s)^4$.}
\label{tab:finite_T_parameters}
\end{table}

\begin{figure}
\begin{center}
\includegraphics*[width=0.75\textwidth]{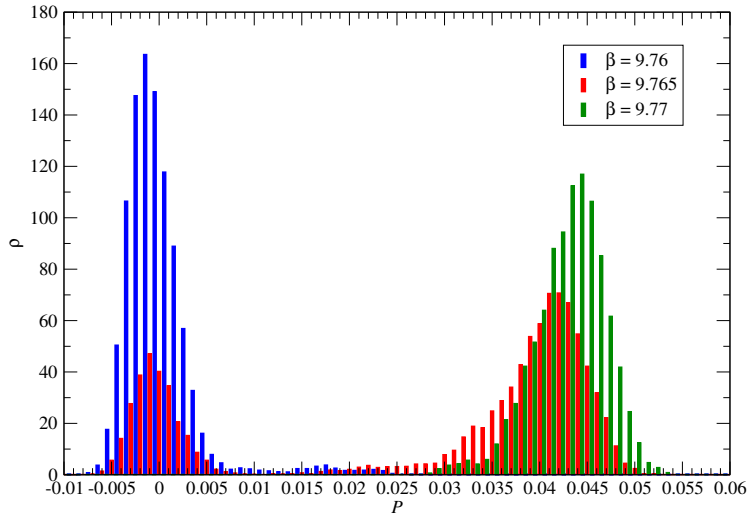}
\caption{The distribution of values for the spatially-averaged Polyakov loop $P$ at three different temperatures, as obtained from simulations on lattices with $N_t=6$ and $N_s=32$ at $\beta=9.76$ (blue histograms), at $\beta=9.765$ (red histograms) and at $\beta=9.77$ (green histograms), reveals the transition from a confining phase at low temperature, in which $\rho(P)$ has a peak close to zero, to a deconfining one at high temperature, in which $\rho$ has a global maximum for a finite value of $P$. 
\label{fig:P_distribution_6x32}}
\end{center}
\end{figure}

The first task consists in identifying, for each value of $N_t$, the critical coupling corresponding to the transition from the confining to the deconfined phase: by varying $\beta$, the lattice spacing $a$ can be tuned to $1/(N_t T_c)$. As mentioned in sect.~\ref{sec:setup}, the transition from one phase to the other can be identified by monitoring how the distribution $\rho(P)$ of values of the spatially averaged, bare Polyakov loop $P$ varies with $\beta$. The confining phase is characterized by a distribution with a peak near zero,\footnote{The fact that, even in the low-temperature regime, the peak is not exactly at zero is related to the absence of an exact center symmetry in this theory, and to the fact that, as a consequence, the trace of the Polyakov loop is not an actual order parameter.} while in the deconfined phase $\rho(P)$ has a maximum at a finite value of $P$, and the transition (or crossover) region can be identified as the one in which $\rho(P)$ takes a double-peak structure, with approximately equal maxima. An example of such behavior is shown in fig.~\ref{fig:P_distribution_6x32}, where we plotted the distribution of values of Polyakov loops from lattices with $N_t=6$ and $N_s=32$, at three different $\beta$ values, namely $9.76$, $9.765$ and $9.77$ (corresponding to three different values of the lattice spacing, and, hence, of the temperature).

More precisely, in any finite-volume lattice this identifies a pseudo-critical coupling: as usual, the existence of a phase transition is only possible for an infinite number of degrees of freedom, namely in the thermodynamic, infinite-volume, limit. Thus, the actual critical point corresponding to the thermodynamic phase transition is obtained by extrapolation of the pseudo-critical couplings to the infinite-volume limit.

A more accurate method to determine the location and nature of the transition is based on the study of the Binder cumulant~\cite{Binder:1981aaa, Challa:1986sk, Lee:1991zz}
\begin{equation}
\label{Binder}
B = 1 - \frac{\langle P^4 \rangle}{3 \langle P^2 \rangle^2},
\end{equation}
which is especially useful in computationally demanding problems (see ref.~\cite[appendix]{deForcrand:2010be} for an example). For a system in the thermodynamic limit, $B$ takes the value zero if the expectation value of $P$ vanishes, while it tends to $2/3$ for a pure phase in which the expectation value of $P$ is finite. Thus, for a second-order phase transition, the values of the Binder cumulant interpolate between these two different limits at ``small'' and ``large'' $\beta$ (i.e. at low and at high temperature, respectively). As the system volume is increased, the Binder cumulant tends to become a function with a sharper and sharper increase in the region corresponding to the critical $\beta$. The critical coupling in the thermodynamic limit can thus be estimated from the crossing of these curves. On the other hand, in the presence of a first-order phase transition, in the thermodynamic limit the Binder cumulant tends to $2/3$ in both phases, whereas it develops a deep minimum near the transition point~\cite{Vollmayr:1993aaa} (see also ref.~\cite{Binder:1992aaa} for a discussion). 

\begin{figure}
\begin{center}
\includegraphics*[width=0.8\textwidth]{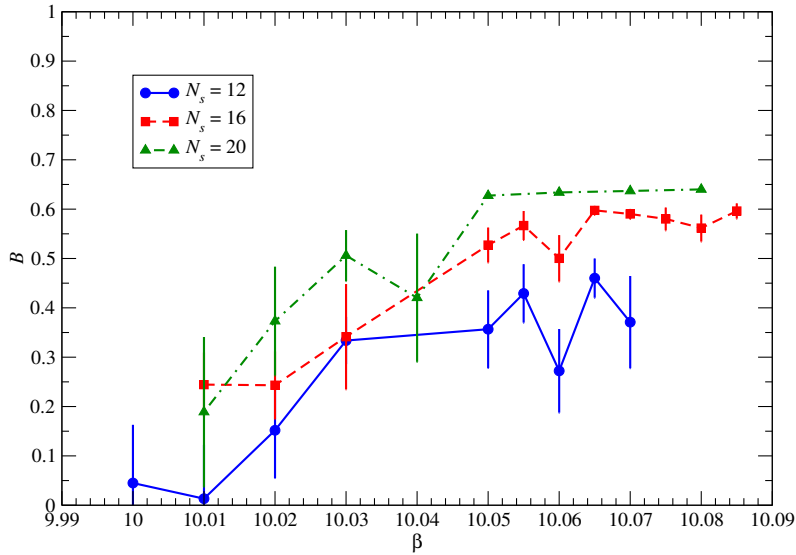}
\caption{The values of the Binder cumulant $B$ defined in eq.~(\protect\ref{Binder}), obtained from lattices of different spatial volume (denoted by symbols of different colors) at fixed $N_t$ (here, $N_t=8$), reveal the location of the (pseudo-)critical $\beta$ at which the deconfinement transition takes place (in this case $10.04 \pm 0.03$).\label{fig:Binder}}
\end{center}
\end{figure}

For our present problem, however, this type of analysis is complicated by the fact that in the thermodynamic limit the expectation value of $P$ in the confining phase is finite, but very small. As a consequence, the behavior that can be observed in numerical simulations for manageable lattice sizes is somewhat different from what one would expect for a first-order phase transition. We studied $B$ for different values of the simulation parameters: one example is shown in figure~\ref{fig:Binder}, which refers to simulations on lattices with $N_t=8$ sites in the Euclidean-time direction, at different values of $\beta$ and for different spatial volumes. Note that, despite the technical challenge that we just mentioned, the values of $B$ vary rapidly within a narrow $\beta$-interval, allowing one to locate the (pseudo-)critical point with quite good precision.

For our present purposes, however, the main qualitative features of the thermal deconfinement transition in $\Gtwo$ Yang-Mills theory are already revealed by how the Monte Carlo history of the spatially averaged Polyakov loop and the $\rho(P)$ distribution vary with $\beta$ and the lattice volume. For $\beta$ values equal to (or larger than) the critical one, the former exhibits tunneling events between different vacua, which become increasingly rare when the lattice volume grows. This suggest that the passage from the confining to the deconfined phase is a transition of first order. Accordingly, the peaks in the $\rho$ distribution at criticality tend to become separated by an interval of $P$ values, whose probability density is exponentially suppressed when the lattice volume increases. 

\begin{figure}
\begin{center}
\includegraphics*[width=0.75\textwidth]{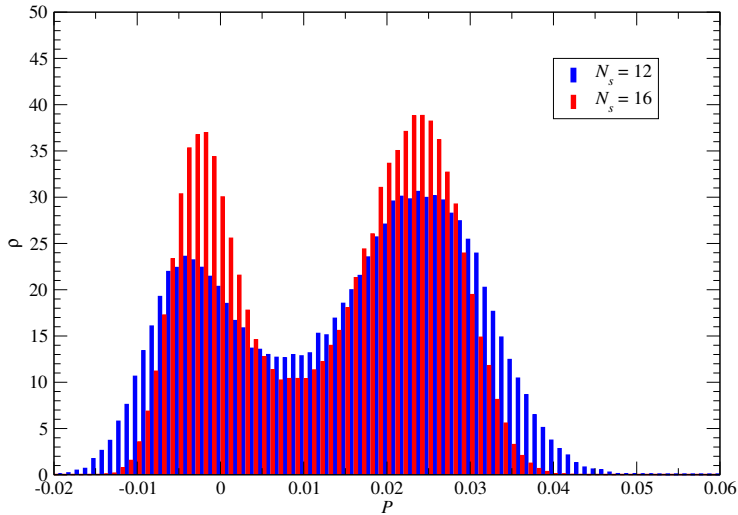}
\caption{Distribution of values of the spatially-averaged Polyakov loop $P$ close to the deconfinement transition, from simulations on lattices of fixed spacing (and Euclidean time extent) and different spatial volumes, corresponding to $N_s=12$ (blue histograms) and $N_s=16$ (red histograms).
\label{fig:P_distribution_critical}}
\end{center}
\end{figure}

As an example, in fig.~\ref{fig:P_distribution_critical} we show the distribution of $P$ values obtained from simulations at the critical point for fixed $N_t=8$ and fixed lattice spacing, for spatial volumes $(12a)^3$ and $(16a)^3$.

Our observation of a first-order deconfining transition confirms the results of earlier lattice studies of this model~\cite{Pepe:2006er, Cossu:2007dk}.

\begin{figure}
\centerline{\includegraphics[width=.75\textwidth]{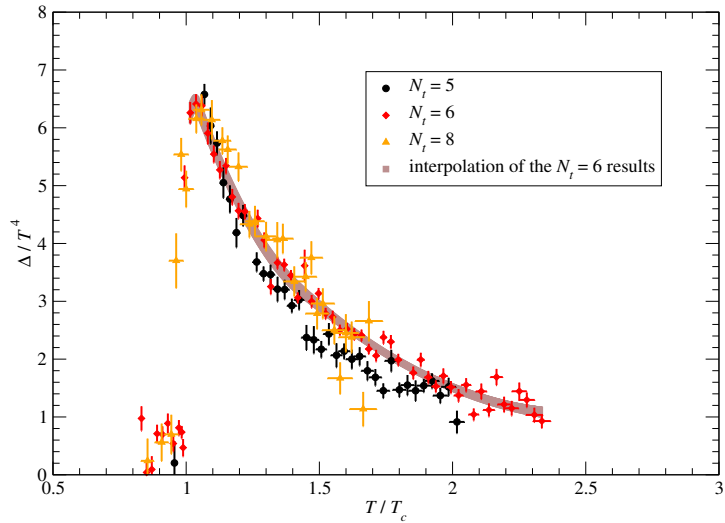}}
\caption{Trace of the energy-momentum tensor, in units of $T^4$, as a function of the temperature $T$ (in units of $T_c$). Results obtained from simulations on lattices with different $N_t$ are denoted by different symbols: black circles for $N_t=5$, red diamonds for $N_t=6$, orange triangles for $N_t=8$. The errorbars shown in the plot account for the statistical uncertainties in the computation of the average plaquettes in eq.~(\protect\ref{lattice_rescaled_trace}), as well as for statistical and systematic uncertainties related to the non-perturbative determination of the scale and of the critical coupling for each $\beta$. The brown band denotes the interpolation of our results from the ensembles corresponding to $N_t=6$. As discussed in the text, such curve turns out to be essentially compatible with the results obtained from an attempt to carry out the continuum extrapolation (up to small deviations in the latter, which are likely due to statistical effects). Thus, we present the brown curve as an approximate estimate of the continuum limit.}
\label{fig:Delta_over_T4}
\end{figure}

Having set the scale and determined the critical coupling for different values of $N_t$, we proceed to the computation of equilibrium thermodynamic quantities at different lattice spacings, and to the discussion of their extrapolation to the continuum limit. As pointed out in sect.~\ref{sec:setup}, the static observables of interest in this work are related to each other by elementary thermodynamic identities. Since our numerical determination of the equation of state is based on the integral method introduced in ref.~\cite{Engels:1990vr}, the quantity which is computed most directly is the trace of the energy-momentum tensor $\Delta$: as shown by eq.~(\ref{lattice_rescaled_trace}), it is just given by the difference between the expectation values of the plaquette at zero and at finite temperature, up to a $\beta$-dependent factor. The results from our simulations (at different values of $N_t$) for the dimensionless $\Delta/T^4$ ratio are shown in fig.~\ref{fig:Delta_over_T4}, as a function of the temperature (in units of the critical temperature). Note that the data from simulation ensembles at different $N_t$ are close to each other, indicating that discretization effects are under good control.

To get results in the continuum limit, we first interpolated our results from each $N_t$ data set with splines, and then tried to carry out an extrapolation to the $a \to 0$ limit, by fitting the values interpolated with the splines (at a sufficiently large number of temperatures) as a function of $1/N_t^2$, including a constant and a linear term. 

The first of these two steps is done as follows: we split the data sets in $\nint$ intervals defined by $\nint-1$ internal knots, and compute $\nint+3$ (or $\nint+2$) basis splines (B-splines): they define a function basis, such that every spline can be written as a linear combination of those. The systematic uncertainties involved in the procedure are related to the choice of the number of knots, and to the spline degree (quadratic or cubic); these uncertainties can be estimated by comparing the $\chi^2$ values obtained for different choices, and turn out not to be large.

As for the second step (the pointwise extrapolation to the continuum limit), however, we observed that it leads to results which are mostly compatible with the curve obtained from the interpolation of the $N_t=6$ data set, except for a few (limited) regions, in which the extrapolation is affected by somewhat larger errorbars---an effect likely due to statistical fluctuations in the ensemble obtained from the finest lattice, which tend to drive the continuum extrapolation. Since the latter effects are obviously unphysical, for the sake of clarity of presentation we decided to consider the curve obtained from interpolation of our $N_t=6$ data (with the associated uncertainties) as an estimate of the continuum limit. This curve corresponds to the brown band plotted in fig.~\ref{fig:Delta_over_T4}. 

Our results for the trace anomaly reveal two very interesting features:
\begin{enumerate}
\item When expressed \emph{per gluon degree of freedom}, i.e. dividing by $2 \times \dadj$ (where $2$ is the number of transverse polarizations for a massless spin-$1$ particle in $3+1$ spacetime dimensions, and $\dadj$ is dimension of the gluon representation, i.e. $\dadj=14$ for $\Gtwo$, and $\dadj=N^2-1$ for $\SU(N)$ gauge group), the results for $\Delta/T^4$ agree with those obtained in $\SU(N)$ Yang-Mills theories.
\item In the deconfined phase (at temperatures up to a few times the deconfinement temperature), $\Delta$ is nearly perfectly proportional to $T^2$.
\end{enumerate}

\begin{figure}
\centerline{\includegraphics[width=.75\textwidth]{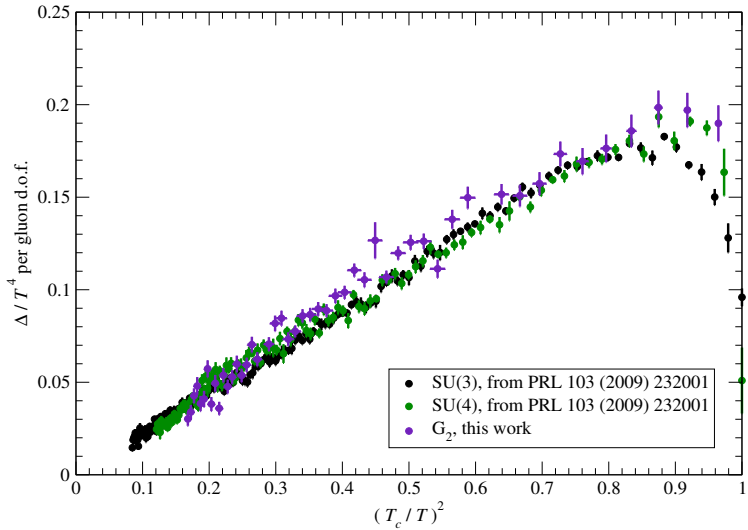}}
\caption{When normalized per gluon d.o.f., the values of $\Delta/T^4$ obtained in $\Gtwo$ Yang-Mills theory (indaco symbols) are compatible with those obtained in theories with a generic $\SU(N)$ gauge group (in particular, here we show the results for $N=3$, denoted by black symbols, and for $N=4$, displayed by green symbols, obtained in ref.~\cite{Panero:2009tv} at $N_t=5$). All $\Gtwo$ results displayed here were obtained from simulations at $N_t=6$. Note that in this figure the data are plotted against $(T_c/T)^2$, in order to reveal the approximately perfect proportionality between $\Delta$ and $T^2$ in the deconfined phase (at least in the temperature range investigated in this work, up to a few times $T_c$): with this choice of axes, this feature manifests itself in the linear behavior observed in the plot.}
\label{fig:Pisarski_per_gluon_dof}
\end{figure}

This is clearly exhibited in fig.~\ref{fig:Pisarski_per_gluon_dof}, where our lattice results for $\Delta/(2 \dadj T^4)$ at $N_t=6$ are plotted against $(T_c/T)^2$, together with analogous results for the $\SU(3)$ and $\SU(4)$ theories (at $N_t=5$) from ref.~\cite{Panero:2009tv}.\footnote{Note that cutoff effects at $N_t \ge 5$ are already rather small, so it is meaningful to compare $N_t=5$ data with those obtained from simulations at $N_t=6$, at least within the precision of our results.} The collapse of data obtained in theories with different gauge groups is manifest, as is the linear dependence on $1/T^2$ in the temperature range shown (implying that $\Delta$ is approximately proportional to $T^2$).

These features were already observed in $\SU(N)$ gauge theories, both in four~\cite{Boyd:1996bx, Bringoltz:2005rr, Panero:2009tv, Datta:2010sq, Borsanyi:2012ve} and in three~\cite{Bialas:2008rk, Caselle:2011mn} spacetime dimensions (the latter provide an interesting theoretical laboratory: see, e.g., ref.~\cite{Frasca:2014uha} and references therein). 

\begin{figure}
\centerline{\includegraphics[width=.75\textwidth]{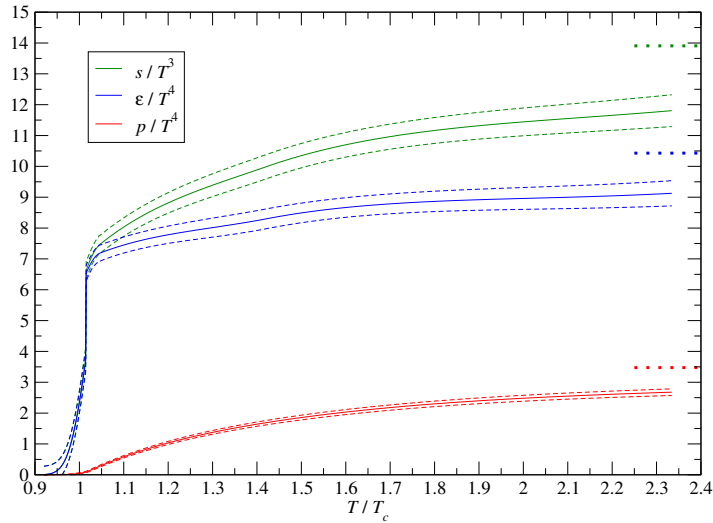}}
\caption{Our estimate for the pressure $p$ in units of $T^4$ (solid red curve), for the energy density $\epsilon$ in units of $T^4$ (solid blue curve) and for the entropy density $s$ in units of $T^3$ (solid green curve), as obtained by numerical integration of our results for $\Delta/T^4$ from the $N_t=6$ ensemble, and according to eq.~(\ref{Delta_epsilon_minus_3p}) and to eq.~(\ref{lattice_entropy_density}), as discussed in the text. The dashed curves indicate the uncertainties affecting each of these observables. These quantities are shown as a function of the temperature $T$, in units of deconfinement temperature $T_c$. The dotted horizontal lines on the right-hand side of the picture represent the values of these observables in the free limit, for the Wilson discretization on a lattice with $N_t=6$.}
\label{fig:p_epsilon_s}
\end{figure}

Integrating the plaquette differences used to evaluate $\Delta/T^4$, the pressure (in units of $T^4$) is then computed according to eq.~(\ref{integral_method}) for each $N_t$. In principle, one could then extrapolate the corresponding results to the continuum limit; however, like for the trace anomaly, it turns out that, at the level of precision of our lattice data, this leads to results which are essentially compatible with those from our $N_t=6$ ensemble (within uncertainties, including those related to the extrapolation systematics). Therefore, in fig.~\ref{fig:p_epsilon_s} we show the results for $p/T^4$ obtained by numerical integration of the curve interpolating the $N_t=6$ data (solid red curve): this curve can be taken as an approximate estimate of the continuum limit (up to an uncertainty defined by the band within the dashed red curves). As one can see, at the highest temperatures probed in this work the pressure is growing very slowly (due to the logarithmic running of the coupling with the typical energy scale of the thermal ensemble, which is of the order of $T$) and tending towards its value in the Stefan-Boltzmann limit\footnote{Strictly speaking, the Stefan-Boltzmann value of $p/T^4$ in the \emph{lattice} theory at finite $N_t$ is different from (in particular: larger than) the continuum one: for $N_t=6$ and for the Wilson gauge action used in this work, the correction is approximately $13\%$. For a detailed derivation, see ref.~\cite[eqs.~(A.5) and~(A.6)]{Caselle:2011mn} and refs.~\cite{Engels:1999tk, Scheideler_thesis}. The horizontal dotted lines on the right-hand side of fig.~\ref{fig:p_epsilon_s}, showing the Stefan-Boltzmann limits for the three observables plotted, take this correction into account.}
\begin{equation}
\label{p_SB}
\left( \frac{p}{T^4} \right)_{\mbox{\tiny{SB}}} = \frac{14}{45} \pi^2,
\end{equation}
so that at temperatures $T \lesssim 3T_c$ the system is still relatively far from a gas of free gluons. Fig.~\ref{fig:p_epsilon_s} also shows our results for the energy density in units of the fourth power of the temperature ($\epsilon/T^4$, solid blue curve) and for the  entropy density in units of the cube of the temperature ($s/T^3$, solid green curve), respectively determined according to eq.~(\ref{Delta_epsilon_minus_3p}) and to eq.~(\ref{lattice_entropy_density}). Like for the pressure, the uncertainties affecting these two quantities are denoted by the bands enclosed by the dashed curves. In the same figure, we also show the values of these quantities in the free limit of the $\Gtwo$ lattice Yang-Mills theory (with the Wilson discretization) for a lattice with $N_t=6$~\cite{Engels:1999tk, Scheideler_thesis, Caselle:2011mn}: these values are denoted by the dotted lines on the right-hand side of the plot.

\section{Discussion}
\label{sec:discussion}

The features of this exceptional group (in particular: the fact that it has a trivial center) make the $\Gtwo$ Yang-Mills model very interesting for a comparison with gauge theories based on classical simple Lie groups. As we discussed, previous works already showed that at zero temperature this model bears several qualitative similarities with QCD: the physical spectrum does not contain colored states, and the potential associated with a pair of static color sources is linearly rising at intermediate distances---before flattening out at very large distances, due to dynamical string-breaking. However, a difference with respect to QCD (in which the color charge is screened by creation of dynamical quark-antiquark pairs, which are absent in pure Yang-Mills theory) is that in $\Gtwo$ Yang-Mills theory screening is due to gluons. At finite temperature, there is numerical evidence that this theory has a first-order deconfinement phase transition (at which the average Polyakov loop modulus jumps from small to finite values), even though this transition is not associated with the breaking of center symmetry~\cite{Pepe:2006er}.

Our lattice results confirm the first-order nature of the deconfinement transition in $\Gtwo$ Yang-Mills theory at finite temperature. This is in agreement with analytical studies available in the literature. In particular, a semiclassical study of the confinement/deconfinement mechanism in different Yang-Mills theories was presented in ref.~\cite{Poppitz:2012nz} (a related study, discussing the inclusion of fundamental fermionic matter, is reported in ref.~\cite{Poppitz:2013zqa}, while a generalization to all simple Lie groups has been recently presented in ref.~\cite{Anber:2014lba}). Generalizing a previous study for the $\SU(2)$ case~\cite{Poppitz:2012sw}, the authors of ref.~~\cite{Poppitz:2012nz} showed how the properties of the high-temperature phase of a generic Yang-Mills theory can be understood, by studying its $\mathcal{N}=1$ supersymmetric counterpart on $\R^3 \times S^1$, and by continuously connecting the supersymmetric model to the pure Yang-Mills theory, via soft supersymmetry-breaking induced by a finite gluino mass. This analytical study is possible, by virtue of the fact that, when the $S^1$ compactification length $L$ becomes small, the theory can be reliably investigated by means of semi-classical methods.\footnote{Note that, since \emph{periodic} boundary conditions are imposed along the compactified direction for all fields (including fermionic ones), the transition in the supersymmetric theory is a \emph{quantum}---rather than a \emph{thermal}---one.} In particular, analyzing the effective potential describing the eigenvalues of the Polyakov line, it turns out that:
\begin{itemize}
\item The phase transition is driven by the competition between terms of perturbative origin~\cite{Gross:1980br, Belyaev:1989yt, Enqvist:1990ae, KorthalsAltes:1993ca}, Bogomol'nyi-Prasad-Sommerfield monopole-instantons, and Kaluza-Klein monopole-instantons (which tend to make the Polyakov-line eigenvalues collapse, namely to break center symmetry) and neutral bions (that stabilize the center).\footnote{For further details about these topological objects, see also refs.~\cite{Unsal:2007jx, Anber:2011de, Argyres:2012ka} and references therein.}
\item This confining/deconfining mechanism is common to all non-Abelian theories, irrespective of the underlying gauge group. The order of the deconfining transition, however, does depend on the gauge group: for the $\SU(2)$ case, the mechanism predicts the existence of a second-order transition, whereas for $\SU(N \ge 3)$ and for $\Gtwo$ the transition is a discontinuous one.
\end{itemize}

Regarding the latter point, it is worth mentioning that in ref.~\cite{Holland:2003kg} it was conjectured that the order of the deconfinement phase transition should be determined by the number of gluon degrees of freedom: for larger Lie groups, the passage from the confining phase (in which the number of hadronic states is independent of the size of the gauge group) to the deconfined phase (in which the number of colored states \emph{does} depend on the group size) corresponds to a more abrupt change in the number of relevant degrees of freedom, that can make the transition of first order. The problem has also been studied in lattice simulations in $2+1$ spacetime dimensions (in which each of the gluon color components has one---rather than two---transverse degree of freedom): there, the deconfinement transition is of second order for $\SU(2)$~\cite{Christensen:1990vc, Teper:1993gp} and $\SU(3)$~\cite{Christensen:1991rx, Bialas:2012qz} Yang-Mills theories (and the critical indices agree with the Svetitsky-Yaffe conjecture~\cite{Svetitsky:1982gs}). For the $\SU(4)$ theory in $2+1$ dimensions, the order of the deconfinement transition is particularly difficult to identify: it has been studied in refs.~\cite{deForcrand:2003wa, Holland:2007ar, Liddle:2008kk}, and the most recent results indicate that the transition is probably a weakly first-order one~\cite{Holland:2007ar, Liddle:2008kk}. For $\SU(5)$ gauge group, the transition is a stronger first-order one~\cite{Holland:2005nd, Liddle:2008kk}, like for $\SU(N>5)$~\cite{Liddle:2008kk}. These results confirm that, like in $3+1$ dimensions, also in $2+1$ dimensions the order of the transition depends on the number of gluon degrees of freedom, with a passage from a continuous to a discontinuous transition when the number of gluon degrees of freedom in the deconfined phase exceeds a number around $15$ (see also ref.~\cite{Buisseret:2011fq} for further comments about this issue); since the number of gluon degrees of freedom in $\Gtwo$ Yang-Mills theory in $2+1$ dimensions is $14$, it would be interesting to investigate whether the transition is of first or of second order.

In addition, the results of our lattice simulations also show that the equilibrium thermodynamic observables in this theory are \emph{quantitatively} very similar to those determined in previous studies of the $\SU(N)$ equation of state~\cite{Bringoltz:2005rr, Panero:2009tv, Datta:2010sq}. In fact, rescaling the various thermodynamic quantities by the number of gluon degrees of freedom, we found that the observables per gluon component in the deconfined phase of $\Gtwo$ Yang-Mills theory are essentially the same as in $\SU(N)$ theories. This is consistent with the observation (based on an analysis of the gluon propagator in Landau gauge) that confinement and deconfinement should not be qualitatively dependent on the gauge group~\cite{Maas:2010qw}. The same independence from the gauge group has also been observed in the numerical study of Polyakov loops in different representations in $\SU(N)$ gauge theories~\cite{Mykkanen:2012ri}: the quantitative similarities with results in the $\SU(3)$ theory~\cite{Kaczmarek:2002mc, Dumitru:2003hp, Gupta:2007ax} are suggestive of common dynamical features.

In particular, our data show that, in the deconfined phase of $\Gtwo$ Yang-Mills theory, the trace of the stress-energy tensor $\Delta$ is nearly exactly proportional to $T^2$ for temperatures up to a few times the critical deconfinement temperature $T_c$. This peculiar behavior was first observed in $\SU(3)$ Yang-Mills theory~\cite{Boyd:1996bx} (see also ref.~\cite{Borsanyi:2012ve} for a more recent, high-precision study) and discussed in refs.~\cite{Meisinger:2001cq, Megias:2005ve, Pisarski:2006yk, Megias:2009mp, Zuo:2014vga}. Later, it was also observed in numerical simulations of gauge theories with $\SU(N>3)$~\cite{Bringoltz:2005rr, Panero:2009tv, Datta:2010sq}. A dependence on the square of the temperature is hard to explain in perturbative terms (unless it accidentally results from cancellations between terms related to different powers of the coupling). Actually, at those, relatively low, temperatures, most likely the gluon plasma is not weakly coupled and non-perturbative effects probably play a non-negligible r\^ole~\cite{Kajantie:2002wa, Hietanen:2008tv}. While one could argue that this numerical evidence in a relatively limited temperature range may not be too compelling, it is interesting to note that lattice results reveal the same $T^2$-dependence also in $2+1$ spacetime dimensions~\cite{Bialas:2008rk, Caselle:2011mn} (for a discussion, see also ref.~\cite{Bicudo:2014cra} and references therein). Note that, in the latter case, due to the dimensionful nature of the gauge coupling $g$, the relation between $g^2$ and the temperature is linear, rather than logarithmic.

Another interesting recent analytical work addressing the thermal properties of $\Gtwo$ Yang-Mills theory was reported in ref.~\cite{Dumitru:2013xna} (see also  ref.~\cite{Guo:2014zra}): following an idea discussed in refs.~\cite{Meisinger:2001cq, Pisarski:2006hz, Dumitru:2010mj, Dumitru:2012fw}, in this article the thermal behavior of the theory near $T_c$ is assumed to depend on a condensate for the Polyakov-line eigenvalues, and the effective action due to quantum fluctuations in the presence of this condensate is computed at two loops. The somewhat surprising result is that the two-loop contribution to the effective action is proportional to the one at one loop: this holds both for $\SU(N)$ and for $\Gtwo$ gauge groups. In order to derive quantitative predictions for the pressure and for the Polyakov loop as a function of the temperature, however, non-perturbative contributions should be included, as discussed in ref.~\cite{Dumitru:2012fw}. More precisely, the effective description of the deconfined phase of Yang-Mills theories presented in ref.~\cite{Dumitru:2012fw} is based on the assumption that, at a given temperature, the system can be modeled by configurations characterized by a constant (i.e. uniform in space) Polyakov line, and that the partition function can be written in terms of an effective potential for the Polyakov-line eigenvalues. By gauge invariance, the Polyakov line can be taken to be a diagonal matrix without loss of generality. For $\SU(N)$ gauge groups, the $N$ eigenvalues of this matrix are complex numbers of modulus $1$. Since the determinant of the matrix equals $1$, the eigenvalues' phases are constrained to sum up to $0 \mod 2 \pi$. It is convenient to define rescaled phases (to be denoted as $\mathbf{q}$), that take values in the real interval between $-1$ and $1$; one can then write an effective potential $\mathcal{V} (\mathbf{q},T)$, which includes different types of contributions (of perturbative and non-perturbative nature). For the $\Gtwo$ gauge group, the construction exploits the fact that $\Gtwo$ is a subgroup of $\SO(7)$, which, in turn, is a subgroup of $\SU(7)$. Starting from $\SU(7)$, these conditions reduce the number of independent components of $\mathbf{q}$ down to $2$, the rank of $\Gtwo$. Thus, the effective potential can take the form
\begin{equation}
\label{tot_potential_generic}
\frac{\mathcal{V} (\mathbf{q},T)}{T^4} = - \frac{14}{45} \pi^2 + \frac{4}{3} \pi^2 V_2(\mathbf{q}) - \frac{4 \pi^2 T_c^2}{3 T^2} \left[ c_1 V_1(\mathbf{q}) + c_2 V_2(\mathbf{q}) + c_3 + d_2 V_2^{(\mathbf{7})}(\mathbf{q}) \right],
\end{equation}
where the first term on the right-hand side simply gives the free-energy density for the free-gluon gas (according to eq.~(\ref{p_SB}) and to the $p=-f$ identity), the next term is the leading perturbative contribution, which can be expressed in terms of Bernoulli polynomials, while the terms within the square brackets are expected to mimic effects relevant close to the deconfinement transition (note the presence of the coefficient proportional to $T_c^2$, related to non-perturbative physics, in front of the square brackets): see ref.~\cite{Dumitru:2012fw} for the precise definitions and for a thorough discussion.

The effective potential in eq.~(\ref{tot_potential_generic}) depends on the unknown coefficients $c_1$, $c_2$, $c_3$ and $d_2$, which, in principle, could be fixed using our data. We carried out a preliminary study in this direction, finding that (with certain, mild assumptions) it is indeed possible to fix values for $c_1$, $c_2$, $c_3$ and $d_2$ yielding $\Delta/T^4$ values compatible with our lattice data. However, the quality of such determination is not very good, because the parameters appear to be cross-correlated and/or poorly constrained. Without imposing additional restrictions, an accurate determination of these parameters would require data of extremely high precision (and a genuine, very accurate continuum extrapolation).

After fixing the parameters of the effective potential defined in eq.~(\ref{tot_potential_generic}), it would be interesting to test, whether the model correctly predicts the behavior of other observables computed on the lattice near the deconfining transition: this is a task that we leave for the future.

\section{Conclusions}
\label{sec:conclusions}

The present lattice study of the $\Gtwo$ Yang-Mills model at finite temperature confirms that this gauge theory has a finite-temperature deconfining phase transition. In agreement with earlier lattice studies~\cite{Pepe:2006er, Cossu:2007dk}, we found that the latter is of first order, as predicted using semi-classical methods applicable to all simple gauge groups~\cite{Poppitz:2012nz}. In particular, the nature of the deconfinement transition, determined by the form of the effective potential experienced by the Polyakov-loop eigenvalues, results from the competition of different topological objects (and perturbative effects~\cite{Gross:1980br}): neutral bions (which generate repulsion among the eigenvalues) and magnetic bions, as well as monopole-instantons (which generate attraction among eigenvalues, like the perturbative terms).

The study of the equation of state that we carried out also shows that the equilibrium thermal properties of $\Gtwo$ gauge theory are qualitatively and quantitatively very similar to those of all $\SU(N)$ theories (up to a trivial proportionality to the number of gluon degrees of freedom), and are compatible with the predictions from recent analytical studies, like those reported in refs.~\cite{Lacroix:2012pt, Dumitru:2012fw, Dumitru:2013xna}. Recently, analogous conclusions have also been reached for supersymmetric theories~\cite{Lacroix:2014gsa, Lacroix:2014jpa}, using an approach inspired by ref.~\cite{Shuryak:2004tx}. These results corroborate the idea of \emph{universality} in the thermal behavior of confining gauge theories. To summarize with a pun, one could say that the exceptional thermodynamics in the title of the present paper ``is not so exceptional, after all''. 

Our findings are also interesting to understand the r\^ole that different topological excitations play in confinement, and give indications about the non-perturbative dynamics relevant at temperatures close to deconfinement, where truncated weak-coupling expansions are no longer quantitatively accurate.

This work could be generalized along various directions. The temperature range that we investigated could be extended, possibly in combination with the technical refinement of using an improved version of the gauge action, as was done for $\SU(3)$ Yang-Mills theory in ref.~\cite{Borsanyi:2012ve}. It is worth remarking that the multilevel algorithm used to set the scale in the present work has been recently generalized to improved actions~\cite{Mykkanen:2012dv}. With sufficient computational power, it would be interesting to compare the behavior of the thermodynamic quantities in the confining phase with a gas of free glueballs, possibly modeling the spectrum of excited states in terms of a vibrating bosonic string. This type of comparison was successfully carried out in ref.~\cite{Meyer:2009tq} for $\SU(3)$ Yang-Mills theory in $3+1$ dimensions and in ref.~\cite{Caselle:2011fy} for $\SU(N)$ theories in $2+1$ dimensions. In fact, the investigation of finite-temperature $\Gtwo$ Yang-Mills theory in $2+1$ dimensions could be another possible generalization of this work: as we already remarked in sect.~\ref{sec:discussion}, the identification of the order of the deconfinement transition in that theory would be particularly interesting.

One could also extend the investigation of the theory, by looking at other observables in the deconfined phase, in particular going beyond those characterizing the equilibrium properties of the QCD plasma. While the present study addresses a model which is interesting as a theoretical test bed, but which is not physically realized in nature, ultimately our aim is to achieve a deeper understanding of phenomenologically relevant aspects of strong interactions at finite temperature. In particular, transport properties describing the real-time evolution of conserved charge densities in the QGP are of the utmost relevance for experimentalists and theorists alike. The lattice investigation of these quantities, however, is particularly challenging (see ref.~\cite{Meyer:2011gj} for a detailed discussion), and until recently has been mostly limited to the $\SU(3)$ theory. Given the aspects of \emph{universality} that seem to emerge from the present study and from previous works, it would be interesting to investigate, whether also the spectral functions related to different transport coefficients in $\Gtwo$ Yang-Mills are qualitatively and quantitatively similar to those extracted in the $\SU(3)$ theory---albeit this may prove computationally very challenging.

Another possible generalization would be to investigate the equation of state in a Yang-Mills theory based on another exceptional gauge group. The ``most natural'' candidate would be the one based on $\Ffour$: this group has rank $4$ and dimension $52$ and, like $\Gtwo$, its center is trivial. The smallest non-trivial irreducible representation of this group, however, is $26$-dimensional, making lattice simulations of this Yang-Mills theory much more demanding from a computational point of view. We are not aware of any previous lattice studies of $\Ffour$ gauge theory.

\acknowledgments
This work is supported by the Spanish MINECO (grant FPA2012-31686 and ``Centro de Excelencia Severo Ochoa'' programme grant SEV-2012-0249). We thank C.~Bonati, F.~Buisseret, P.~de~Forcrand, A.~Dumitru, J.~Greensite, K.~Holland, C.~Korthals~Altes, M.~Pepe, R.~Pisarski and U.-J.~Wiese for helpful discussions and comments. The simulations were performed on the PAX cluster at DESY, Zeuthen.

\appendix
\section{General properties of the $\Gtwo$ group and of its algebra}
\label{app:G2_properties}

In this appendix we summarize some basic facts about the $\Gtwo$ group and its algebra. Our discussion mostly follows ref.~\cite{Holland:2003jy}, although (where appropriate) we also provide some additional technical details---in particular as it concerns the representation theory.

$\Gtwo$ is the smallest of the five exceptional simple Lie groups, with $\dimension  \Gtwo =  14$. It is a subgroup of $\SO(7)$ and coincides with the automorphism group of the octonions. Equivalently, it can be defined as the subgroup of $\GL(7)$ preserving the canonical differential $3$-form (given by the canonical bilinear form taking the cross product of two vectors as its second argument). $\Gtwo$ has an $\SU(3)$ subgroup, and $\Gtwo / \SU(3)$ is isomorphic to the six-dimensional sphere $S^6$~\cite{Macfarlane:2002hr}. This allows one to decompose a generic $\Gtwo$ element as the product of a matrix associated with an element of $S^6$, times an $\SU(3)$ matrix: for an explicit realization, see ref.~\cite[appendix A]{Pepe:2006er}. Another subgroup of $\Gtwo$ is $\SO(4)$~\cite{Yokota_0902.0431}.

The Lie algebra of the $\Gtwo$ generators has dimension $14$ and rank $2$: its Cartan matrix is
\begin{equation}
\label{Cartan_matrix}
\left[
\begin{array}{cc}
  2 & -3 \\
 -1 &  2 \\
\end{array}
\right],
\end{equation}
so that the $\Pi$-system includes a long and a short root, at a relative angle $5 \pi/6$ (a unique property among all simple Lie algebras). There exist two fundamental representations, which are seven- and fourteen-dimensional, respectively. The weight diagram of the $7$-dimensional fundamental representation is given by the vertices of a regular hexagon, plus its center. The representation of dimension $14$ is the adjoint representation: its weight diagram is given by the vertices of a hexagram, with the addition of two points at its center. 

The irreducible representations can be unambiguously labeled by two non-negative integers, $(\lambda_1,\lambda_2)$; all of the irreducible representations can be cast in real form. The dimension ($d$) of a generic irreducible representation of label $(\lambda_1,\lambda_2)$ is given by the Weyl dimension formula
\eq
\label{dimensions}
d = \frac{(2 \lambda_1+3 \lambda_2 +5) \cdot (\lambda_2 +1) \cdot (\lambda_1+3 \lambda_2 +4) \cdot (\lambda_1+ \lambda_2 +2) \cdot (\lambda_1+1) \cdot (\lambda_1+2\lambda_2+3)}{120}.
\en
The trivial representation corresponds to $(0,0)$, whereas the fundamental representation of dimension $7$ is associated with labels $(1,0)$, while the adjoint corresponds to $(0,1)$. As a curiosity, note that the dimension of the representation $(9,9)$ is exactly one million.
All irreducible representations of dimension not larger than $10^5$ are listed in table~\ref{irreps}.

\begin{table*}[h!]
\centering
\phantom{-------}
\begin{tabular}{||c|c||c|c||c|c||c|c||}
\hline
 $(\lambda_1,\lambda_2)$ & $d$ & $(\lambda_1,\lambda_2)$ & $d$ & $(\lambda_1,\lambda_2)$ & $d$ & $(\lambda_1,\lambda_2)$ & $d$ \\    
\hline
  $(0,0)$ & $\mathbf{1}$  & $(6,1)$ & $\mathbf{3003}$ & $(3,5)$ & $\mathbf{18304}$  & $(4,6)$ & $\mathbf{53599}$ \\
  $(1,0)$ & $\mathbf{7}$  & $(9,0)$ & $\mathbf{3289}$ & $(6,3)$ & $\mathbf{19019}$  & $(7,4)$ & $\mathbf{55614}$ \\
  $(0,1)$ & $\mathbf{14}$ & $(0,6)$ & $\mathbf{3542}$ & $(0,9)$ & $\mathbf{19096}$  & $(2,8)$ & $\mathbf{56133}$ \\
  $(2,0)$ & $\mathbf{27}$ & $(3,3)$ & $\mathbf{4096}$ & $(2,6)$ & $\mathbf{19278}$  & $(11,2)$ & $\mathbf{56133^{\prime}}$ \\
  $(1,1)$ & $\mathbf{64}$ & $(2,4)$ & $\mathbf{4914}$ & $(8,2)$ & $\mathbf{19683}$  & $(3,7)$ & $\mathbf{57344}$ \\
  $(0,2)$ & $\mathbf{77}$ & $(1,5)$ & $\mathbf{4928}$ & $(14,0)$ & $\mathbf{20196}$ & $(9,3)$ & $\mathbf{59136}$ \\
  $(3,0)$ & $\mathbf{77^{\prime}}$  & $(7,1)$ & $\mathbf{4928^{\prime}}$ & $(11,1)$ & $\mathbf{24192}$ & $(18,0)$ & $\mathbf{59983}$ \\
  $(4,0)$ & $\mathbf{182}$ & $(10,0)$ & $\mathbf{5005}$ & $(5,4)$ & $\mathbf{24948}$  & $(14,1)$ & $\mathbf{61047}$ \\
  $(2,1)$ & $\mathbf{189}$ & $(5,2)$ & $\mathbf{5103}$  & $(15,0)$ & $\mathbf{27132}$ & $(0,12)$ & $\mathbf{67158}$ \\
  $(0,3)$ & $\mathbf{273}$ & $(0,7)$ & $\mathbf{6630}$  & $(9,2)$ & $\mathbf{28652}$  & $(6,5)$ & $\mathbf{69160}$ \\
  $(1,2)$ & $\mathbf{286}$ & $(4,3)$ & $\mathbf{7293}$  & $(7,3)$ & $\mathbf{28672}$  & $(1,10)$ & $\mathbf{74074}$ \\
  $(5,0)$ & $\mathbf{378}$ & $(11,0)$ & $\mathbf{7371}$ & $(1,8)$ & $\mathbf{29667}$  & $(12,2)$ & $\mathbf{76076}$ \\
  $(3,1)$ & $\mathbf{448}$ & $(8,1)$ & $\mathbf{7722}$  & $(0,10)$ & $\mathbf{30107}$ & $(19,0)$ & $\mathbf{76153}$ \\
  $(6,0)$ & $\mathbf{714}$ & $(6,2)$ & $\mathbf{8372}$  & $(4,5)$ & $\mathbf{30107^{\prime}}$ & $(8,4)$ & $\mathbf{79002}$ \\
  $(2,2)$ & $\mathbf{729}$ & $(3,4)$ & $\mathbf{9177}$  & $(3,6)$ & $\mathbf{33495}$  & $(15,1)$ & $\mathbf{80256}$ \\
  $(0,4)$ & $\mathbf{748}$ & $(1,6)$ & $\mathbf{9660}$  & $(12,1)$ & $\mathbf{33592}$ & $(5,6)$ & $\mathbf{81081}$ \\
  $(1,3)$ & $\mathbf{896}$ & $(2,5)$ & $\mathbf{10206}$ & $(2,7)$ & $\mathbf{33858}$  & $(10,3)$ & $\mathbf{81719}$ \\
  $(4,1)$ & $\mathbf{924}$ & $(12,0)$ & $\mathbf{10556}$& $(16,0)$ & $\mathbf{35853}$ & $(2,9)$ & $\mathbf{88803}$ \\
  $(7,0)$ & $\mathbf{1254}$& $(0,8)$ & $\mathbf{11571}$ & $(6,4)$ & $\mathbf{37961}$  & $(4,7)$ & $\mathbf{89726}$ \\
  $(3,2)$ & $\mathbf{1547}$& $(9,1)$ & $\mathbf{11648}$ & $(10,2)$ & $\mathbf{40579}$ & $(3,8)$ & $\mathbf{93093}$ \\
  $(5,1)$ & $\mathbf{1728}$& $(5,3)$ & $\mathbf{12096}$ & $(8,3)$ & $\mathbf{41769}$  & $(20,0)$ & $\mathbf{95634}$ \\
  $(0,5)$ & $\mathbf{1729}$& $(7,2)$ & $\mathbf{13090}$ & $(0,11)$ & $\mathbf{45695}$ & $(0,13)$ & $\mathbf{96019}$ \\
  $(2,3)$ & $\mathbf{2079}$& $(13,0)$ & $\mathbf{14756}$& $(13,1)$ & $\mathbf{45696}$ & $(7,5)$ & $\mathbf{99008}$ \\
  $(8,0)$ & $\mathbf{2079^{\prime}}$& $(4,4)$ & $\mathbf{15625}$ & $(5,5)$ & $\mathbf{46656}$  & & \\
  $(1,4)$ & $\mathbf{2261}$& $(10,1)$ & $\mathbf{17017}$& $(17,0)$ & $\mathbf{46683}$ & & \\
  $(4,2)$ & $\mathbf{2926}$& $(1,7)$ & $\mathbf{17472}$ & $(1,9)$ & $\mathbf{47872}$  & & \\
\hline
\end{tabular}
\phantom{-------}
\caption{The smallest irreducible representations of the $\Gtwo$ group, sorted by increasing dimension $d$, up to $10^5$. $\mathbf{1}$ denotes the trivial representation, while $\mathbf{7}$ and $\mathbf{14}$ denote the two fundamental representations ($\mathbf{14}$ being the adjoint representation). Non-equivalent irreducible representations of the same dimension are distinguished by a prime sign (we conventionally choose to use the prime sign for the representation with the largest value of $\lambda_1$). Note that, in contrast to the claim of ref.~\cite{Wikipedia_representations}, there exists only one irreducible representation of dimension $28652$.}
\label{irreps}
\end{table*}

Tensor products of irreducible representations are not, in general, irreducible. However, they can be decomposed into sums of irreducible representations. For $\Gtwo$, the most straightforward algorithm to compute the decomposition of tensor products of irreducible representations is the one based on girdles (see ref.~\cite{Behrends:1962zz} and references therein), which can be briefly summarized as follows.

In the real vector space of dimension equal to the rank $n$ of the group ($\R^2$ in this case) with coordinates $x_1, \dots , x_n$, consider $t$ distinct points $P^{(i)}=\left( P^{(i)}_1, \dots, P^{(i)}_n \right)$, with $i=1, \dots , t$, and define the corresponding set of points  $\mathcal{S}$ by assigning integer multiplicities $k^{(i)}$ to each $P^{(i)}$:
\begin{equation}
\label{point_set}
\mathcal{S}=\left\{ \left( P^{(1)} ; k^{(1)} \right), \dots , \left( P^{(t)} ; k^{(t)} \right) \right\}.
\end{equation}
A generic set of points $\mathcal{S}$ can then be uniquely associated with a Laurent polynomial in $n$ variables
\eq
\label{Laurent_polynomial}
\sigma( x_1, \dots x_n ) = \sum_{i=1}^t k^{(i)} \left( \prod_{j=1}^{n} x_j^{P^{(i)}_j} \right),
\en
and the operations of addition, subtraction, multiplication and division of sets of points are then defined by the result of the same operations on the associated polynomials.

The girdle $\xi(\lambda_1, \dots \lambda_n)$ of a representation of a group is then a particular set of points, with certain well-defined multiplicities: for $\Gtwo$, the girdles are irregular dodecagons, that are symmetric with respect to both the $x_1$ and $x_2$ axes, and whose vertices in the first quadrant are listed in table~\ref{girdle_points_I_quadrant}. The multiplicities associated with the vertices in the other quadrants are also $\pm 1$, and are alternating around the dodecagon.\footnote{Thus, for example, for the trivial representation the point of coordinates $x_1=-1/\left(4\sqrt{3}\right)$, $x_2=3/4$ has multiplicity $-1$, while the one of coordinates $x_1=-1/\sqrt{3}$, $x_2=1/2$ has multiplicity $1$, and so on.}

\begin{table*}[h!]
\centering
\phantom{-------}
\begin{tabular}{||c||c|c|c||}
\hline
$i$ & $P^{(i)}_1$ & $P^{(i)}_2$ & $k^{(i)}$ \\
\hline
$1$ & $(2\lambda_1+3\lambda_2+5)/\sqrt{48}$ & $(\lambda_2+1)/4$ & $1$ \\
$2$ & $(\lambda_1+3\lambda_2+4)/\sqrt{48}$ & $(\lambda_1+\lambda_2+2)/4$ & $-1$ \\
$3$ & $(\lambda_1+1)/\sqrt{48}$ & $(\lambda_1+2\lambda_2+3)/4$ & $1$ \\
\hline
\end{tabular}
\phantom{-------}
\caption{Coordinates and multiplicity coefficients of the points belonging to the girdle associated with the $\Gtwo$ representation of label $(\lambda_1,\lambda_2)$, within the first quadrant of the $\R^2$ plane. Note that, in each coordinate, the constant term equals the sum of the coefficients of $\lambda_1$ and $\lambda_2$.}
\label{girdle_points_I_quadrant}
\end{table*}

The character of a given representation with label $(\lambda_1,\lambda_2)$ is given by the ratio of polynomials of two girdles:
\eq
\chi(\lambda_1,\lambda_2) = \frac{\xi(\lambda_1,\lambda_2)}{\xi(0,0)}.
\label{character}
\en
This allows one to decompose arbitrary tensor products of irreducible representations using the fact that, since
\eq
\chi(\lambda_1,\lambda_2) \chi(\mu_1,\mu_2) = \sum_{(\nu_1,\nu_2)} q_{(\nu_1,\nu_2)} \chi(\nu_1,\nu_2),
\label{character_decomposition}
\en
one also has
\eq
\frac{\xi(\lambda_1,\lambda_2) \xi(\mu_1,\mu_2)}{\xi(0,0)} = \sum_{(\nu_1,\nu_2)} q_{(\nu_1,\nu_2)} \xi(\nu_1,\nu_2),
\label{girdle_decomposition}
\en
which immediately allows one to identify the $q_{(\nu_1,\nu_2)}$ coefficients,
since only girdles appear on the right-hand side of eq.~(\ref{girdle_decomposition}).

A numerical implementation of the algorithm described above immediately shows that, in particular, the following decomposition laws for tensor products of $\Gtwo$ representations hold:
\eqar
\mathbf{7} \otimes \mathbf{7} &=& \mathbf{1} \oplus \mathbf{7} \oplus \mathbf{14} \oplus \mathbf{27}, \label{7_otimes_7} \\
\mathbf{14} \otimes \mathbf{7} &=& \mathbf{7} \oplus \mathbf{27} \oplus \mathbf{64}, \label{14_otimes_7} \\
\mathbf{27} \otimes \mathbf{7} &=& \mathbf{7} \oplus \mathbf{14} \oplus \mathbf{27} \oplus \mathbf{64} \oplus \mathbf{77}, \label{27_otimes_7} \\
\mathbf{14} \otimes \mathbf{14} &=& \mathbf{1} \oplus \mathbf{14} \oplus \mathbf{27} \oplus \mathbf{77} \oplus \mathbf{77^\prime}, \label{14_otimes_14} \\
\mathbf{27} \otimes \mathbf{14} &=& \mathbf{7} \oplus \mathbf{14} \oplus \mathbf{27} \oplus \mathbf{64} \oplus \mathbf{77} \oplus \mathbf{189}, \label{27_otimes_14} \\
\mathbf{77} \otimes \mathbf{14} &=& \mathbf{14} \oplus \mathbf{27} \oplus \mathbf{64} \oplus \mathbf{77} \oplus \mathbf{77^\prime} \oplus \mathbf{182} \oplus \mathbf{189} \oplus \mathbf{448}, \label{77_otimes_14} \\
\mathbf{77^\prime} \otimes \mathbf{14} &=& \mathbf{14} \oplus \mathbf{77} \oplus \mathbf{77^\prime} \oplus \mathbf{189} \oplus \mathbf{273} \oplus \mathbf{448}. \label{77prime_otimes_14}
\enar
Note that, since both the trivial ($\mathbf{1}$) \emph{and} the smallest fundamental ($\mathbf{7}$) representation appear on the r.h.s. of eq.~(\ref{7_otimes_7}), it is elementary to prove by induction that (at least in principle) $\Gtwo$ QCD admits color-singlet ``hadrons'' made of any number $\nval \ge 2$ of valence quarks: ``diquarks'', ``baryons'', ``tetraquarks'', ``pentaquarks'', ``hexaquarks'', ``heptaquarks'', et~c.

Using the formulas above, the tensor product of three adjoint representations $\mathbf{14}$ can be decomposed as
\eqar
\label{screening}
\mathbf{14} \otimes \mathbf{14} \otimes \mathbf{14} &=& \mathbf{1} \oplus \mathbf{7} \oplus \mathbf{14} \oplus \mathbf{14} \oplus \mathbf{14} \oplus \mathbf{14} \oplus \mathbf{14} \oplus \mathbf{27} \oplus \mathbf{27} \oplus \mathbf{27} \oplus \mathbf{64} \oplus \mathbf{64} \nonumber \\
& & \oplus \mathbf{77} \oplus \mathbf{77} \oplus \mathbf{77} \oplus \mathbf{77} \oplus \mathbf{77^\prime} \oplus \mathbf{77^\prime} \oplus \mathbf{77^\prime} \oplus \mathbf{182} \oplus \mathbf{189} \oplus \mathbf{189} \oplus \mathbf{189} \nonumber \\
& &  \oplus \mathbf{273} \oplus \mathbf{448} \oplus \mathbf{448}.
\enar
The presence of the representation $\mathbf{7}$ on the right-hand side of eq.~(\ref{screening}) implies that in $\Gtwo$ Yang-Mills theory a fundamental color source can be screened by three gluons.

The $\Gtwo$ Casimir operators are discussed in ref.~\cite{Bincer:1993jb}; in particular, the functionally independent ones are those of degree $2$ and $6$. Their eigenvalues can be found in ref.~\cite[section 5]{Englefield:1980ih}.

The non-perturbative aspects of a non-Abelian gauge theory are related to the topological objects that gauge field configurations can sustain. For the $\Gtwo$ group, the homotopy groups are listed in table~\ref{homotopy_groups}---see ref.~\cite{Mimura1967}---, where $\Z_1$ denotes the trivial group. $\Gtwo$ is connected, with a trivial fundamental group; its second homotopy group is trivial, too, while the third is the group of integers, hence $\Gtwo$ gauge theory admits ``instantons''.

\begin{table*}[h!]
\centering
\phantom{-------}
\begin{tabular}{||c|c||c|c||c|c||}
\hline
 $n$ & $\pi_n(\Gtwo)$ & $n$ & $\pi_n(\Gtwo)$ & $n$ & $\pi_n(\Gtwo)$ \\
 \hline
 $0$ & $\Z_1$ & $6$  & $\Z_3$           & $12$ & $\Z_1$                 \\
 $1$ & $\Z_1$ & $7$  & $\Z_1$           & $13$ & $\Z_1$                 \\
 $2$ & $\Z_1$ & $8$  & $\Z_2$           & $14$ & $\Z_{168} \oplus \Z_2$ \\
 $3$ &  $\Z$  & $9$  & $\Z_6$           & $15$ & $\Z_6 \oplus \Z_2 \oplus \Z_2$ \\
 $4$ & $\Z_1$ & $10$ & $\Z_1$           & $16$ & $\Z_8 \oplus \Z_2$      \\
 $5$ & $\Z_1$ & $11$ & $\Z \oplus \Z_2$ & $17$ & $\Z_{240}$              \\
\hline
\end{tabular}
\phantom{-------}
\caption{Lowest homotopy groups of the $\Gtwo$ group, from ref.~\cite{Mimura1967}.}
\label{homotopy_groups}
\end{table*}

Finally, using the properties of exact sequences (and the additivity of homotopy groups of direct products of groups), it is also trivial to show that $\pi_2\left( \Gtwo / [ \U(1) \times \U(1) ] \right) $ is $\Z \oplus \Z$: this implies that, like for $\SU(3)$ gauge theory, when the global $\Gtwo$ group gets broken to its Cartan subgroup, two types of 't~Hooft-Polyakov monopoles appear.

\bibliography{paper}

\end{document}